\documentclass[12pt, a4paper]{report}
\usepackage[left=1.5cm, top=2cm, right=1.5cm, bottom=2cm]{geometry}
\usepackage[utf8]{inputenc}
\usepackage{graphicx}
\usepackage{amsmath}
\usepackage{subfigure}

\begin{document}
\thispagestyle{empty}
\begin{center}

{\LARGE \textbf{\textsf{\textsl{Cerebral Synchrony Assessment Tutorial: A General Review on Cerebral Signals' Synchronization Estimation Concepts and Methods\footnote[1]{\textsl{Version \#3.2}}}}}}

\vspace{1.5in}

by

\vspace{0.3in}
{\textbf{{\Large Esmaeil Seraj$ ^{\dagger, *} $ \footnote[2]{\textsl{Email: esmaeil.seraj09@gmail.com}}}}}
\vspace{1in}

\emph{ $ ^\dagger $Signal Processing Center, Department of Computer Science and Engineering and Information Technology, School of Electrical and Computer Engineering, Shiraz University, Shiraz, Iran\footnote[3]{\textsl{Previous affiliation}}}

\vspace{0.1in}
\emph{ $ ^* $School of Electrical and Computer Engineering, Georgia Institute of Technology, Atlanta, Georgia, United States\footnote[4]{\textsl{Current affiliation}}}

\vspace{2in}
\textbf{July 2018\footnote[5]{\textsl{Last update}}}
\end{center}
\newpage
\pagenumbering{roman}
\begin{abstract}
The human brain is ultimately responsible for all thoughts and movements that the body produces. This allows humans to successfully interact with their environment. If the brain is not functioning properly many abilities of human can be damaged. The goal of cerebral signal analysis is to learn about brain function. 

The idea that distinct areas of the brain are responsible for specific tasks, the functional segregation, is a key aspect of brain function. Functional integration is an important feature of brain function, it is the concordance of multiple segregated brain areas to produce a unified response. There is an amplified feedback mechanism in the brain called reentry which requires specific timing relations. This specific timing requires neurons within an assembly to synchronize their firing rates. This has led to increased interest and use of phase variables, particularly their synchronization, to measure connectivity in cerebral signals. Herein, we propose a comprehensive review on concepts and methods previously presented for assessing cerebral synchrony, with focus on phase synchronization, as a tool for brain connectivity evaluation.

\vspace{1cm}
\end{abstract}

\tableofcontents

\newpage
\pagenumbering{arabic}
\chapter{Introduction}
Cerebral signals are among the easiest, most useful and efficient tools for studying brain and its different cognitive states. According to their concept (i.e. embedded information) and also the recording procedure, these signals are called in many different names such as EEG (electroencephalogram), EMG (magnetoencephalogram), ERP (event related potential) or etc.. Analyzing these signals needs some powerful and discriminative  features extracted from them. Amplitude is one of these powerful features. Amplitude of Biological signals has been shown that is very informative and is used as main analysis factor in former studies.

In some applications, because of the environment being too much noisy, or presence of other biological contaminants, the amplitude information of captured signals become contaminated and inadequate. As an example of such applications we can refer to ERP signals studies in BCI purposes that the spontaneous EEG and other biological signals such as EOG play the contaminant role and make the amplitude of ERP signals inadequate in information. As an other instance, investigating fetal brain signals needs recordings from maternal abdominal. Also in this application there are some biological contaminants such as fetal and maternal cardiac activity signals (fECG, mECG). These signals are stronger, i.e. 10--100 fold, in amplitude than fetal brain signals. Thus, the amplitude solely can not yield adequate information. In such cases we need to look for new features to use and capture information from new aspects that are less contaminated and noisy.

Phase of this biological signals is an informative useful alternative feature in these cases that has been of interest in past few years. In many recent researches it has been shown that phase has valuable information and can be used as a complement for amplitude information. Thus, using phase of brain signals in cases that amplitude informations are contaminated or inadequate, or the phase itself has adjunct information over amplitude (i.e. brain cognitive response investigations), would be very helpful.

Phase of cerebral signals has been extracted and used through various procedures in past studies such as \emph{Fourier} and \emph{Wavelet} transforms in frequency domain, \emph{Analytic} form of signals and also using \emph{multi--channel} signals and the angles between signal subspaces. Each of these methods are applied on data sets related to a specific application such as brain computer interface purposes, cognitive researches, event related potentials studies and etc. Since for different applications the result of all these various methods will vary, the important confronting problem here is that there is not a comprehensive and reliable comparison between these methods. So we decided to first specify these different methods and extract a well-defined conceptual framework out of them to prevent from being perplexed in using phase analysis methods. After that we will implement some of these methods on specific data sets in specific applications to compare the results and specify the properties of each method in details. We hope that the results of this study will be useful in applications such as ERP signals analysis (i.e. BCI systems studies) and also brain cognitive responses investigations.

To this goal, in Chapter 1, the brain signals' phase analysis methods from two viewpoints are presented: (1) the early analysis and (2) the modern analysis. Different related quantities are defined and the statistics are presented and some benefits and limitations confronting in both the early and the modern viewpoints are discussed. Finally at the end of this section the related applications are introduced. Phase analysis is of an extended interest in this applications because of their particular characteristics.

In Chapter 2, phase analysis methods are presented and investigated afterward in different fields such as measuring connectivity in distinct brain areas, phase modulations and demodulations and high order statistics. The different approaches in each specific case such as \emph{Coherence}, \emph{Phase Synchronization } and \emph{Desynchronization} and etc. are presented in this Section. The statistics and the motivation of each method are proposed and discussed for each part.



\section{Cerebral Signals: Early Analysis}
The human brain is ultimately responsible for all thought and movement that the body produces. This allows humans to successfully interact with their environment. If the brain is not functioning properly many abilities of human can be damaged. The goal of cerebral signal analysis is to learn about brain function.

Cerebral signals are susceptible to be contaminated by various noise sources such as muscular activities (i.e. blinking, jaw clenching or frowning), artifact generated by the electrical equipments which are near recording areas and also other biological signals (i.e. EMG or EOG). Artifact detection and removal is a necessary step for many types of cerebral signal analysis [1].

Early cerebral signal recording systems were very simple in comparison to the their modern species. The analogue EEG signals recorded at that time were analyzed visually and the qualitative features of them were been investigated usually. Features of the EEG signal were classified as either paroxysmal (transient, bursting) or on-going (background, spontaneous) activity. A simple feature of EEG recordings was measured by counting the number of oscillations in one second.

Because of recordings being too much noisy the quantification and interpretation of the amplitude of a desired frequency component were too difficult, since the peak-to-peak amplitude of the signals changed over time. One approximation used was to create an envelope for the signal by joining together all the peaks/troughs and taking an average of the resulting lines [2]. Another used measure was the duration of time that a recording would spend oscillating in a particular frequency band [3]. 

Many complex waveform that exist in EEG recording such as the alpha rhythm which was the first EEG waveform reported by Hans Berger [4] (also called as the Berger waveform) and the K-complex which is a transient waveform that occurs during sleep were discovered [5](many more examples can be found in [6]). Also topographic maps of the scalp potentials were used to visually identify spatial patterns in EEG signals, however creating these maps were difficult from analogue recordings.

Often visual analysis was not sufficient to discriminate between components and furthermore, more complex events were very cumbersome with analogue recordings. Analogue-to-digital converters (ADC) are used to create digital EEG recordings. Digital EEG recordings opened new possibilities for analyzing the spectral content of brain signals. Creating topographic maps was made more practical with digital recordings and systems. Using the digital EEGs, computers used interpolation to create detailed contour plots of the scalp potential. However, there are also some ambiguities remaining when interpolation is used, due to projecting a 3D volume onto a 2D surface [7]. 

Based on the idea of \emph{Fourier analysis}, using the Fourier transform leads to measuring the \emph{power spectral density}(PSD) and so the \emph{mean squared coherence} (MSC) (which employs the PSD) and also extracting significant frequency specific components such as instantaneous phase and amplitude.

\subsubsection{Power Spectral Density}
Let  $ x(t) $ be a stochastic stationary process, The Fourier transform of $ x(t) $ is shown by $ X(f) $ and is determined as below:
\begin{equation}
X(f)=\int_{-\infty}^{\infty}x(t)e^{-j2\pi ft}dt
\end{equation}
Then the power spectral density $ PSD_x(f) $ of $ x(t) $ is calculated as follow:
\begin{equation}
PSD_x(f)=\sum_{i=-\infty}^{\infty}|X_i(f)|^2
\end{equation}
The PSD can also be calculated between two distinct signals and it is called as cross-spectrum function. Cross-spectrum is a complex function that its amplitude and phase show the correlation between two signals. For instance, the cross-spectrum of two sinusoidal signals of the same frequency has a sharp peak in that frequency. The phase of the cross-spectrum at that frequency equals the phase difference between the records. If two signals share a sinusoidal activity, but each also contains other, unshared activity, their cross-spectrum has a peak at the shared frequency only.

\subsubsection{Correlation Function}
Another measure availed through digital signal recordings is the correlation function. Auto-correlation function $ R_x(\tau) $ of $ x(t) $ is measured as below:
\begin{equation}
R_x(\tau)=x(\tau)x(-\tau)=\int_{-\infty}^{\infty}x(t)x(t+\tau)dt
\end{equation}
The Fourier transform of the Auto-correlation function does indeed equal to the $ PSD $ of our stochastic stationary signal $ x(t) $.
\begin{equation}
PSD_x(f)=\int_{-\infty}^{+\infty}R_x(\tau)e^{j2\pi f\tau}d\tau
\end{equation}
The proof is quite straightforward, by mean of Fourier transform of the correlation function $ R_x(\tau) $ we have:
$$
f\{R_x(\tau)\}=\int_{-\infty}^{\infty}R(\tau)e^{-j2\pi f\tau}d\tau
$$
$$
=\int_{-\infty}^{\infty}\int_{-\infty}^{\infty}x(t)x(t+\tau)e^{-j2\pi f\tau}dtd\tau
$$
$$
=\int_{-\infty}^{\infty}x(t)\int_{-\infty}^{\infty}x(t+\tau)e^{-j2\pi f\tau}d\tau dt
$$
$$
=X(f)\int_{-\infty}^{\infty}x(t)e^{j2\pi ft}dt
$$
$$
=X(f)X^*(f)=|X(f)|^2
$$
\subsubsection{Phase \& Amplitude Extraction}
Two other quantities availed through digitalized early cerebral signal analysis are the instantaneous phase and amplitude. There are two conceptually distinct set of techniques for the purpose of capturing the amplitude and phase of a signal: (1) linear techniques and (2) nonlinear techniques. In one hand, the linear techniques assume a constant amplitude and phase within the estimation window [8]. As a common example, Fourier transform represents the stationary process $ x(t) $ in the frequency space as a phasor $ X(f) $:
$$
X(f)=\int_{-\infty}^{\infty}x(t)e^{-j2\pi ft}dt=A(f)\cos\theta(f)+jA(f)\sin\theta(f)=A(f)e^{j\theta(f)}
$$
where $ A(f) $ and $ \theta(f) $ are the amplitude and phase of signal $ x(t) $ within the estimation window respectively.

In the other hand, nonlinear techniques try to measure the time-dependent instantaneous amplitude and phase which are illustrating the moment-to-moment change in signal. Instantaneous amplitude and phase are usually obtained using either the \emph{Hilbert} or \emph{Wavelet} transforms which has been illustrated that both produce similar results [9]. There are other recently proposed methods for phase extraction purpose which are represented here.

\paragraph{Hilbert Transform}
The Hilbert transform of a real valued process $ x(t) $ is defined as:
\begin{equation}
x_h(t)=\frac{1}{\pi}PV\int_{-\infty}^{\infty}\frac{x(\tau)}{t-\tau}d\tau
\end{equation}
where $ x_h(t)=H\{x(t)\} $ represents the Hilbert transform of process $ x(t) $ and the integral is taken using the Cauchy principal value due to the potential singularity at $ t=\tau $. The Hilbert transform is the convolution of the signal $ x(t) $ with the function $ \frac{1}{\pi t} $ which means the PSD of $ x_h(t) $ is equal to the product of the PSD of $ x(t) $ and the PSD of $ \frac{1}{\pi t} $. Besides, the Fourier transform of the function $ \frac{1}{\pi t} $ is equal to:
$$
F\{\frac{1}{\pi t}\}=-isgn(f)
$$
which states that the Hilbert transform is simply a $ \frac{\pi}{2} $ shift in the phase of the original signal. Thus using the analytic extension of a signal as presented below does not change the PSD of that signal [10]:
\begin{equation}
Z(t)=x(t)+jx_h(t)
\end{equation}
Now, the instantaneous amplitude and phase of $ x(t) $ can be defined as the magnitude and argument of the analytic extension as follow:
\begin{equation}
Z(t)=A(t)e^{j\phi(t)}
\end{equation}
where $ A(t) $ is the instantaneous amplitude and $ \phi(t) $ is the instantaneous phase which can be calculated as below:
\begin{equation}
\phi(t)=\arctan\left(\frac{x_h(t)}{x(t)}\right)
\end{equation}
To have a physically meaningful interpretation of the instantaneous phase and amplitude, they have to be extracted from a narrow-band signal or in a specific frequency of interest, so a filtering pre-step before applying Hilbert transform is needed [11].

\paragraph{Wavelet Transform}
As mentioned before, another useful procedure for measuring the instantaneous amplitude and phase of a signal is through time-frequency transformations such as Wavelet transform. By mean of the standard Morlet-Wavelet, the Wavelet coefficients of signal $ x(t) $ can be determined as follow [12]:
\begin{equation}
W_x(t,f)=\int_{-\infty}^{\infty}x(u)\psi_{t,f}^*(u)du
\end{equation}
where the $ \psi_{t,f}(u) $ is a modulated Gaussian function (with center of $ t $ and variance of $ \sigma $) by a sinusoidal wave with frequency of $ f $:
$$
\psi_{t,f}(u)=\sqrt{f}e^{j2\pi f(u-t)}e^{-\frac{(u-t)^2}{2\sigma^2}}
$$
Then two important parameters obtained from Wavelet transform are \emph{spectrogram} and \emph{Instantaneous Phase} [12]. We have:
\begin{equation}
Spectrogram=|W_x(t,f)|^2
\end{equation}
and
\begin{equation}
exp(j\phi_x(t,f))=\frac{W_x(t,f)}{|W_x(t,f)|}
\end{equation}
which in latter the $ \phi_x(t,f) $ represents the instantaneous phase.

\paragraph{Phase Extraction Based on Complex Energy Density Function}
\textquotedblleft Rihaczek derived the signal energy distribution in time and frequency by application of the complex signal notation\textquotedblright [71].

Considering two complex signals $ x(t) $ and $ y(t) $ which are at similar frequency, the total complex energy can be defined as:
\begin{equation}
CE(t,f)=\int_{-\infty}^{\infty} x(t)y^*(t)dt
\end{equation}

Rihaczek utilized this idea to compute the interaction energy of a signal within some frequency band. This leads to the \emph{Complex Energy Density Function} (CEDF), stated as below [72]:
\begin{equation}
CEDF(t,f)=\frac{1}{\sqrt{2\pi}}x(t)X^*(f)\exp(-j2\pi ft)
\end{equation}

CEDF measures the complex energy of a signal around time $ t $ and frequency $ f $. As stated in [71], CEDF provides a better understanding of phase-modulated signals. Now, the time-varying phase for signal $ x(t) $ in time-frequency plane is defined as below:
\begin{equation}
\phi_x(t,f)=\arg\left(\frac{CEDF(t,f)}{|CEDF(t,f)|}\right)
\end{equation}

\paragraph{Phase Extraction Based on Empirical Mode Decomposition} 
The \emph{Empirical Mode Decomposition} EMD technique was first proposed by Huang in 1998 [76], and is based on the assumption that any signal consists of different simple intrinsic independent modes of oscillation. Each of these modes will have the same number of extrema and zero-crossings and there is only one extremum between successive zero-crossings [74]. Here we present the statistics given in [76] for EMD algorithm with respect to the signal $ x(t) $:
\begin{itemize}
\item  Find all the local \emph{extrema}. Then connect all the local maxima together and also all the local minima by a cubic spline as the \emph{upper and lower envelope} respectively, $ (e_{max}(t) and e_{min}(t)) $.
\item Extract the \emph{detail} as the difference between the signal and the mean of the upper and lower envelope values : $ d(t)=x(t)-\frac{e_{max}(t)+e_{min}(t)}{2} $.
\item If $ d(t) $ is not an IMF, treat it as original signal and repeat first two steps until $ d(t) $ becomes an IMF, then $ d(t)=I_1(t) $ is the first IMF.
\item After getting the first IMF, remove it from the original signal and obtain the residual $ r_1 $: $ x(t)-I_1(t)=r_1(t) $.
\item Treat the $ r_1(t) $ as the original signal and repeat the above steps to obtain the second IMF, $ I_2(t) $.
\item Repeat the process $ n $ times. The decomposition process can be stopped when no more IMF components can be extracted from the last residual.
\end{itemize}
Thus, the EMD algorithm decomposes the signal $ x(t) $ into $ n $ empirical modes as:
\begin{equation}
x(t)=\sum_{i=1}^{n}I_i(t)+r_n(t)
\end{equation}

The IMFs $ I_1(t),...,I_n(t) $, include different frequency bands ranging from high to low [74]. Now, after decomposing signal $ x(t) $ into its relative IMFs, for the purpose of phase extraction, Hilbert transform can be applied to the IMFs and produce the instantaneous phase sequences.

The main difference between this method and other phase extraction method is that, here, phase sequences are extracted from IMFs and there is no need for band-pass filtering the signal anymore. Also EMD based methods have some other merits which are discussed in next chapter while introducing EMD based phase analysis method for assessing synchronization in brain.

\subsubsection{Coherence}
The idea that distinct areas of the brain are responsible for specific tasks, the functional segregation, is a key aspect of brain function. Functional integration is another important feature of brain function, it is the concordance of multiple segregated brain areas to produce a unified response [13], [14].

To measure functional integration in brain, different procedures were introduced to quantify functional connectivity, a statistical dependence between different cerebral signals captured from different brain regions. The first famous measure of functional connectivity was coherence.

Coherence determines the correlation between the signals at specific frequencies [15], [16]. Coherence in a simple definition is a measure of synchronization and correlation between two random process, stochastic, wide-sense stationary (\emph{WSS}) signals that is computable through the coherence function. Coherence is widely used to study dependency and relationship between different brain regions particularly during a specific cognitive task or while experiencing a specific stimuli [17]. 

In general Coherence function is a complex valued number with both an amplitude and phase that are used to measure phase synchronization in signals. The main advantage of coherence over other correlation calculating methods (i.e. in [18]) is that in fact coherence gives the correlation between two signals as a function of frequency that provides the possibility of studying spatial correlation in different frequency bands [17]. 

According to the different characteristics of signals (i.e. stochastic or deterministic) and also different procedures presented for determining PSDs in literature, several approaches for MSC calculation exist. Here we represent the most widely used approaches and in next Chapter after reintroducing the conventional MSC, The rest of the approaches will be discussed.

\paragraph{MSC Using FFT}
In [17] a method for measuring coherence function is presented. The method is based on weighted windowing of the Fourier transform of signals. Let $ x(t) $ and $ y(t) $ be two random process, zero mean, wide--sense stationary and ergodic signals with length $ l $. A summary of the method in [17] is given below:
\begin{itemize}
\item Divide the signals into $ N $ equal parts with $ T $ samples (different part can be overlapped or disjoint)
\item Multiply the samples of each ensemble in a weighting function (i.e. cosine)
\item Take \emph{FFT} of each weighted ensemble
\item Measure power spectral densities of signals (\emph{PSD})
\item Finally, measure the coherence function (or as said \textquotedblleft\emph{Mean Squared Coherence (MSC)\textquotedblright})
\end{itemize}

If $ X_i(f) $ and $ Y_i(f) $ be the Fourier transform of i-th ensemble, the PSDs are calculated as follow:
\begin{equation}
PSD_{xx}(f)=\frac{1}{NT}\sum_{i=1}^{N}|X_i(f)|^2
\end{equation}
\begin{equation}
PSD_{yy}(f)=\frac{1}{NT}\sum_{i=1}^{N}|Y_i(f)|^2
\end{equation}
\begin{equation}
PSD_{xy}(f)=\frac{1}{NT}\sum_{i=1}^{N}X_i(f)Y_i^*(f)
\end{equation}

With the ergodicity of signals being neglected we have:
\begin{equation}
PSD_{xx}(f)=E\{|X_i(f)|^2\}
\end{equation}
\begin{equation}
PSD_{yy}(f)=E\{|Y_i(f)|^2\}
\end{equation}
\begin{equation}
PSD_{xy}(f)=E\{X_i(f)Y_i^*(f)\}
\end{equation}  
       
where $ E\{.\} $ illustrates the mathematical expectation. Using these PSDs we compute the mean squared coherence \emph{MSC} as:
\begin{equation}
|MSC(f)|^2=\frac{|PSD_{xy}(f)|^2}{PSD_{xx}(f)PSD_{yy}(f)}=\frac{\frac{1}{N}\sum_{i=1}^{N}|(PSD_{xy}(f))_i|^2}{PSD_{xx}(f)PSD_{yy}(f)}
\end{equation}

Since $ PSD_{xy}(f) $ is complex, the \emph{Cartesian} or \emph{Polar} coordinates can be used for averaging. We discussed This issue in Appendix A.

\paragraph{MSC Using Correlation Functions}
Another method for obtaining complex coherence is presented in [20] which computes the PSDs using auto and cross correlations. In this method after dividing the signals into $ N $ separate parts, first the auto/cross correlations $ R_{xx}, R_{yy} $ and $ R_{xy} $ are calculated and then the PSDs are measured by taking Fourier transform of these correlation functions:
\begin{equation}
R_{xx}(\tau)=E\{x(t)x(t+\tau)\}\Rightarrow PSD_{xx}=\int_{-\infty}^{+\infty}R_{xx}(\tau)e^{j2\pi f\tau}d\tau
\end{equation}
\begin{equation}
R_{yy}(\tau)=E\{y(t)y(t+\tau)\}\Rightarrow PSD_{yy}=\int_{-\infty}^{+\infty}R_{yy}(\tau)e^{j2\pi f\tau}d\tau
\end{equation}
\begin{equation}
R_{xy}(\tau)=E\{x(t)y(t+\tau)\}\Rightarrow PSD_{xy}=\int_{-\infty}^{+\infty}R_{xy}(\tau)e^{j2\pi f\tau}d\tau
\end{equation} 
                          
The equations (1.23), (1.24) and (1.25) can also be illustrated using inner product by the complex sinusoidal mother function as follow:
\begin{equation}
PSD_{xx}=<R_{xx},e^{j2\pi ft}>
\end{equation}
\begin{equation}
PSD_{yy}=<R_{yy},e^{j2\pi ft}>
\end{equation}
\begin{equation}
PSD_{xy}=<R_{xy},e^{j2\pi ft}>
\end{equation}

This method also takes all the assumptions considered in former method. Here, as well, if the ergodicity assumption is neglected thus the expectation term in correlation calculation could be changed by time averaging. Finally the MSC is obtained as below:
\begin{equation}
MSC(f)=\frac{PSD_{xy}(f)}{\sqrt{PSD_{xx}(f)PSD_{yy}(f)}}=\frac{F\{R_{xy(f)}\}}{\sqrt{F\{R_{xx}(f)\}F\{R_{yy}(f)\}}}
\end{equation}
where $ F\{.\} $ denotes the Fourier transform.

MSC varies between 0 to 1 ($0<MSC<1$). The more two signals are related and synchronized, the grater the MSC value will become. In fact a MSC value close to 1 means the activity between corresponding signals follows a linear transformation in that frequency of interest and a MSC near zero shows two non-related signals in that frequency.

As discussed before, cross-spectrum is a complex function that its amplitude and phase show the correlation between two signals. For instance, the cross-spectrum of two sinusoidal signals of the same frequency has a sharp peak in that frequency. The phase of the cross-spectrum at that frequency equals the phase difference between the records. If two signals share a sinusoidal activity, but each also contains other, unshared activity, their cross-spectrum has a peak at the shared frequency only. The MSC has the same interpretation to the cross-spectrum and concluded from the discussion above, it is said that the phase of MSC is meaningful and equals the phase difference between two corresponding signals, if the amplitude in that frequency is considerable. The large value of coherence amplitude in a phase of 0 degree indicates the large positive correlation and in a phase of 180 degree it's the vise versa and the the rest of the phase values are interpreted similarly due to the amplitude in that frequency [18]. 

The discriminant factor between methods presented in [17] and [20] is the procedure of measuring power spectral densities of signals. As reported in [20], the PSD can be estimated using non-parametric methods such as Blackman--Tukey method, weighted overlapped segment averaging (\emph{WOSA}), also known as Welch's method, Lag-reshaping method or parametric ones, as discussed later in next Chapter.

\section{Cerebral Signal: Modern Analysis-Cerebral Connectivity}
Brain connectivity depicts patterns of links in the brain. Most of the brain functions, for instance the ones involved in learning, memory, behavior adaptation to stimuli, emotions, as well as pathological processes in some brain disorders like epilepsy, autism or schizophrenia, are based on interactions between neuronal assemblies distributed within and across distinct cerebral regions [21]. For instance, it has been shown that segregated areas may activate in response to a particular cognitive task. Neural assemblies are the basic concept of current scientific model for how cerebral functional integration is achieved. Integration of cerebral areas can be measured by assessing brain connectivity [21], [22], [23]. Brain connectivity can be separated into three different concepts: (1) \emph{structural} (or \emph{neuroanatomical}) connectivity, which denotes anatomical links, (2) \emph{functional} connectivity, which rely on statistical dependencies between signals from different brain areas and (3) \emph{effective} connectivity, which introduces causal interactions among these signals [21], [23], [24]. 

There are reasons discussed in [25] which make \emph{structural connectivity} difficult to define intrinsically. \emph{functional connectivity} is defined as the temporal correlation among the activity of different neural assemblies whereas the direct or indirect effect that one neural system exerts over another is defined as \emph{effective connectivity} [26], [27]. 

Functional and effective connectivity techniques are intensely dependent on measuring the correspondence of neural signals over time. Thus cerebral signal recording techniques such as EEG and MEG with good temporal resolution, are optimal for calculating such connectivity. Most frequently used measures for connectivity are: correlation, coherence, mutual information, transfer entropy, generalized synchronization [28], continuity measure [29], synchronization likelihood [30] and phase synchronization [31].

There is an amplified feedback mechanism called reentry which requires specific timing relations. This specific timing requires neurons within an assembly to synchronize their firing rates. This has led to increased interest and use of phase variables, particularly their synchronization, to measure connectivity in cerebral signals [16]. Moreover, our focus in this work is based on signal phase and its analysis methods. Thus, we will introduce the phase synchrony as a subset of modern methods for cerebral signal analysis and discuss various approaches presented to achieve this quantity in next section.

The next section is dedicated to introducing synchronization and its different aspects. Each case is introduced and its characteristics are represented briefly. Finally the phase synchronization as one of the major investigated fields in this thesis and its various approaches are presented.

\section{Synchronization}
Connectivity measures are widely used to determine the level of synchronization between distinct brain regions using cerebral recordings. As discussed before, since \emph{structural connectivity} is difficult to define inherently, only the two remaining connectivity measures (\emph{functional connectivity} and \emph{effective connectivity}) are used for determining synchronization in brain.

As discussed in [23], methods to measure \emph{effective connectivity} can be subdivided into two main categories: (1) model-based and (2) data-driven techniques. The former case assumes theoretical models that describe how brain areas interact and influence each other, whereas the latter case does not assume any specific underlying model or prior knowledge concerning spatial or temporal relationships. Granger-causality (GC), directed transfer function (DTF), directed coherence (DC) and its extension partial directed coherence (PDC) are the most common data-driven connectivity techniques. It is notable that DTF, DC and PDC are developed out of the GC method. The PDC technique will be presented in next chapter as an extension of coherence methods.

The \emph{functional connectivity} methods are subdivided into three main categories: (1) linear, (2) nonlinear and (3) information-based techniques [23]. Each of these categories and the corresponding most common approaches are introduced in next subsections.

\subsection{Linear Synchronization Methods}
\subsubsection{Cross Correlation}
Cross-correlation and MSC are the most commonly used linear synchronization methods. As we discussed before (see Sections 1.2.1.2 and 1.2.1.4), assuming two discrete time processes $ x_n $ and $ y_n $ where $ n=1,...,N $, then the cross-correlation function $ R_{xy} $ is defined as:
\begin{equation}
R_{xy}(\tau)=\frac{1}{N-\tau}\sum_{n=1}^{N-\tau}(\frac{x_n-\mu_x}{\sigma_x})(\frac{y_{n+\tau}-\mu_y}{\sigma_y})
\end{equation}
where $ \mu $ and $ \sigma $ denote mean and variance, respectively, while $ \tau $ is the time lag.

\subsubsection{MSC}
Another linear technique for synchronization measurements is the MSC or simply coherence. MSC is defined precisely in Section (1.1.0.4). To remind what the basic idea is, the main formula to calculate MSC is given below:
\begin{equation}
|MSC(f)|^2=\frac{\frac{1}{N}\sum_{i=1}^{N}|(PSD_{xy}(f))_i|^2}{PSD_{xx}(f)PSD_{yy}(f)}
\end{equation}
which is averaged over trials.

Although MSC is a successful measure of functional integration, it has several limitations [16]:
\begin{itemize}
\item Stationarity: Because of using FFT, existence of a PSD depends on the stationarity of a process. As reported in [32], there is evidence that EEG recordings longer than a particular period of time are non-stationary.
\item Linearity: Coherence is based on linear correlation in the process. Thus, if information is being transferred in a nonlinear manner, it may not be detected.
\item Direction: When there is synchronization between two signals, coherence does not clarify which signal is driving the flow of information.
\end{itemize}

\subsection{Nonlinear Synchronization Methods}
Many crucial neural processes have nonlinear characteristics. Due to the lack of linear synchronization methods to detect this nonlinear connectivity, basic motivation for developing nonlinear methods were not put on outperforming linear methods but rather providing complementary information under certain assumptions [23]. 

Within the field of neuroscience, synchronization is commonly illustrated in the concept of phase synchrony. There are various approaches to obtain this quantity based on different characterization procedures. The main ideas are described briefly in next subsection and more comprehensively in next chapter. Another nonlinear connectivity measure termed as \emph{Generalized synchronization} is also introduced very briefly.

\subsubsection{Phase Synchronization}
Phase synchronization (PS) is defined as temporal adjustment of rhythms of a pair of signals in a frequency band of interest, whereas the amplitudes can remain uncorrelated [33]. PS also can be determined through successive frequency bands to provide a complete measure of synchrony between two signals. Phase locking-value (PLV) is the most commonly used measure of PS. In order to detect periods of phase synchrony (i.e. PLV) in cerebral signals, there are four main steps summarized as follow [34], [35], [36], [37], [38]:
\begin{itemize}
\item The first step is to calculate a sequence consisting instantaneous frequency specific phase values within a temporal window.
\item The second step simply is to determine the instantaneous phase-differences (in frequency $ f $ and time $ t $) from the corresponding phase sequences captured in first step.
\item Quantifying the local stability of this phase-differences across trials of cerebral signals.
\item Determining the degree of statistical significance of each quantity obtained in step three.
\end{itemize}
The first step can be performed through one of the two common signal phase extraction methods: (1) Hilbert Transform or (2) Wavelet transform (see Section 1.2.1.3). The second step actually depends on the third step. The quantification in step three can be performed through various statistical dependence parameters such as mean phase difference, circular variance, standard deviation, Shannon entropy, or mutual information [33]. Depending on which of these procedures to be selected, the process in second step might be required or not. If the phase differences have to be found, there are different ways to this goal depending on the method used for phase extraction. This might differ between simply subtracting the corresponding time-dependent phases captured through \emph{Hilbert Transform} or using description below due to phase extraction through \emph{Wavelet Transform}:
\begin{equation}
\exp(j(\phi_y(t,f)-\phi_x(t,f)))=\frac{W_x(t,f)W_y^*(t,f)}{|W_x(t,f)||W_y(t,f)|}
\end{equation}
where the term $ \phi_y(t,f)-\phi_x(t,f) $ represents the phase-difference between processes $ y $ and $ x $. The main conventional approaches to assess the synchrony between neural signals are distinct in their technique to quantify the captured instantaneous phase-differences. Finally, for the last step commonly surrogate data are utilized. There are different techniques presented for generating surrogate data in literature such as shifting or scrambling the original series [35], [37]. The degree of statistical significance of the PLVs is then determined by comparing them to values obtained from surrogate data.

Generally speaking, for two neural signals $ x(t) $ and $ y(t) $ and their corresponding phases $ \phi_x(t) $ and $ \phi_y(t) $, phase synchrony in its most general form is defined as:
\begin{equation}
n\phi_x(t)-m\phi_y(t)=const.
\end{equation}
where $ n$ and $ m $ are integers indicating the ratios of possible frequency locking [31]. In what follows we assume $ n=m=1 $ for simplicity. Because of problems such as \emph{Volume Conduction} (introduced later in current chapter) or \emph{background spontaneous brain signals} the true synchrony is buried in a considerable background noise [38]. In this case, as in [31], this condition of phase locking can be replaced by the weaker condition as stated below:
\begin{equation}
|n\phi_x(t)-m\phi_y(t)|<const.
\end{equation}

As discussed in [33], such approaches to phase synchrony impose two main limitations as presented below:
\begin{itemize}
\item As mentioned, only synchrony between pairs of signals can be directly studied. This extremely increases the computational cost. For instance, when there is $ n $ signals, $ d=\frac{n^2-n}{2} $ pairs of signal exist.
\item As illustrated in [33], synchronizations with frequency non-stationarity can be observed in brain signals (frequency of
synchronization alters continuously through time, although the phase-difference remains stable). Thus, defining phase synchronization in this manner will prevent such phase synchronizations from being detected.
\item It has been demonstrated in [40] that synchrony-measurement methods based on this definition may be insensitive to very short periods of phase synchronization (see also [41]).
\end{itemize}

The relation between coherence and phase synchronization is investigated in [60] and according to the reported results there is a very close relationship between these concepts. In many studies it has been reported that a complete phase synchronization is manifested by highly coherent phases and correlated amplitudes. In fact, both the phase synchronization and the coherence analysis are looking for periods of phase synchrony in various frequency bands but there is some differences between their sensitivity to detecting them. With focusing on Signals phase, the coherence analysis has some objections (see Section 1.3.1.2), however, in [60] it is reported that phase synchronization analysis is not superior than coherence considerably.

\subsubsection{Generalized synchronization}
After the successful application of PS in EEG analysis, the concept of Generalized synchronization (GS) was introduced [42]. GS approaches were proposed to investigate the dependencies between nonlinear signals without any knowledge about the governing equations [22]. GS represents how much neighborhoods of one chaotic attractor can be mapped onto the other [23]. Although this mapping is prone to stationarity deficits, it is considered to be a robust way of evaluating the GS [42]. Generating such attractors needs delay vectors to be constructed from time series using a procedure known as time-delay embedding [43].

\subsection{Information-based Techniques}
In probability theory and information theory, Mutual Information (MI) measures the mutual dependencies between two random-variable processes by quantifying the amount of information gained about one signal from measuring the other. If two signals have a statistical dependence, observing one signal gains information about the other and reduces the entropy because of the knowledge of other signal. The benefit of using information-based techniques is that these techniques are sensitive to both linear and nonlinear statistical dependencies between signals [16], [23]. 

MI analysis can be used to assess functional connectivity between cerebral signals. For this purpose a measure of entropy is needed [44]. Defining the entropy for a signal $ x(t) $ as:
\begin{equation}
H(x)=-\int_x P(x)\log(P(x))
\end{equation}
we have the mutual information between signals $ x(t) $ and $ y(t) $ determined as:
\begin{equation}
I(x,y)=\int_y\int_xP(x,y)\log(\frac{P(x,y)}{P(x)P(y)})dxdy
\end{equation}
where $ P(x,y) $ is the joint probability density function of $ x(t) $ and $ y(t) $, and $ P(x) $ and $ P(y) $ are the marginal probability density functions of $ x(t) $ and $ y(t) $ respectively [44]. MI has been used for evaluating functional connectivity in EEG recordings (see [45] and [46]). Also MI and entropy based features have been used previously for CAP detection in sleep EEG recordings \cite{ESeraj8}.

\section{Cerebral Signal: Modern Analysis-Higher order Spectra}
The previous introduced statistical tools utilize first and second order statistics to extract information from random signals. However, the main problem confronts in the presence of nonlinearity in systems where the first and second order statistics are unable to adequately analyze many signals [78]. Presenting \emph{High-order Spectra (Statistics)} (HOS) is mainly motivated to overcome this problem and plays an important role in digital signal processing. HOS methods are very useful in problems where non-Gaussian, non-minimum phase, phase coupling or nonlinear behavior and robustness to additive noise are important [78]. They are also used in detection and classification in communication and pattern recognition applications.

\textquotedblleft In general, there are three motivations behind the use of HOS in signal processing which can be considered as its property over lower order spectra: (1) suppressing Gaussian noise of unknown mean and variance; (2) detecting and characterize nonlinearities in the data and (3) reconstructing and preserving the phase as well as the magnitude response of signals or systems\textquotedblright [79]. Also the applications of HOS on biomedical signals are based on these properties. Particularly in this contribution we focus on the latter property of HOS. In fact, measuring Power spectrum causes phase relations and signals phase informations to be suppressed, while HOS,  based on the ability of cumulant spectra to preserve the Fourier-phase of signals, contains such informations [80], [81] and [82]. 

Higher order spectra are functions of two or more component frequencies while the power spectrum and lower order spectra are functions of a single frequency [78]. The \emph{bispectrum} and \emph{bicoherence} and their related extensions (i.e. auto/cross bispectrum and bicoherence) are the third order spectra and also the most used HOS quantities as they are easiest computationally. The other less common HOS quantities than the 3-rd orders are the (auto/cross) \emph{trispectrum} and \emph{tricoherence}. First, we need to review approaches for calculating moments and cumulants of order $ n $ as high order spectra quantities' computation is based on them. 

\subsection{Moments and Cumulants of order $ n $}
For a stationary discrete time random process $ X(k) $, the moments of order $ n $ are given by:
\begin{equation}
m_n(\tau_1,\tau_2,...,\tau_{n-1})=E\{X(k)X(k+\tau_1)...X(k+\tau_{n-1})\}
\end{equation}
where $ E\{.\} $ denotes expectation.

The $ n-th $ order cumulants are functions of the moments of order up to $ n $, for example the first order cumulants can be obtained as below:
\begin{equation}
c_1=m_1=E\{X(k)\}
\end{equation}
which is the \emph{mean}. Consequently, the second order cumulants are given by:
\begin{equation}
c_2(\tau_1)=m_2(\tau_1)-(m_1)^2
\end{equation}
which is the \emph{covariance}. Also for the third order cumulants we have:
\begin{equation}
c_3(\tau_1,\tau_2)=m_3(\tau_1,\tau_2)-(m_1)[m_2(\tau_1)+m_2(\tau_2)+m_2(\tau_2-\tau_1)]+2(m_1)^3
\end{equation}
where $ m_3(\tau_1,\tau_2) $ in the third order moment sequence. Following this manner, the $ n-th $ order cumulants can be obtained. The general relations between cumulants and moments are given in [81].

\subsection{Bispectrum}
\subsubsection{Auto-bispectrum}
The bispectrum of a signal $ x(t) $ can be defined in two ways. First is \emph{the \textbf{triple} product of the DFT's} [81]:
\begin{equation}
B_{xxx}(f_1,f_2)=\lim\limits_{T\rightarrow\infty}\frac{1}{T}E\{X_T(f_1)X_T(f_2)X_T^*(f_1+f_2)\}
\end{equation}

The bispectrum is complex. It means that it contains Fourier magnitude and phase information [83]. With the above definition we have:
\begin{equation}
B_{xxx}(f_1,f_2)=|B_xxx(f_1,f_2)|\exp(j\phi(f_1,f_2))
\end{equation}
where the magnitude of bispectrum obtains as:
\begin{equation}
|B_{xxx}(f_1,f_2)|=|X(f_1)||X(f_2)||X^*(f_1+f_2)|
\end{equation}
and the phase of bispectrum (or biphase [83]) is:
\begin{equation}
\phi(f_1,f_2)=\phi(f_1)+\phi(f_2)-\phi(f_1+f_2)
\end{equation}

Second definition of bispectrum is \emph{the \textbf{double} DFT of the 3rd order cumulant}. Two methods for calculating bispectrum based on this definition in real data are presented later in current section.

The auto-bispectrum is a function of two frequency components, $ f_1 $ and $ f_2 $ which would have a small value if the biphase varies over the different realizations and conversely, would have a large value if this phase does not vary, which is an indication of quadratic coupling between $ f_1 $ and $ f_2 $ [84].

\subsubsection{Cross-bispectrum}
Following the definition of the auto-bispectrum, the cross-bispectrum is used to determine the quadratic coupling between frequency components in two different signals. Considering two signals with Fourier transforms as $ X(f) $ and $ Y(f) $, the cross-bispectrum is defined by [84]:
\begin{equation}
B_{xxy}(f_1,f_2)=\lim\limits_{T\rightarrow\infty}\frac{1}{T}E\{X_T(f_1)X_T(f_2)Y_T^*(f_1+f_2)\}
\end{equation}

Through this definition, the coupling level between two frequency components in $ X(f) $ namely $ f_1 $ and $ f_2 $ and their algebraic sum in $ Y(f) $ is determined which is based on the phase relation between these components in the different realizations. A large value of the cross-bispectrum indicates high \emph{Quadratic Phase Coupling} (QPC) between $ X(f) $ and $ Y(f) $. In fact QPC normally is obtained through the normalized extension of bispectrum known as bicoherence. Bicoherence as a tool for phase analysis is presented within next Chapter [84].

\subsubsection{Methods for Calculating Bispectrum}
There are two main methods for calculating bispectrum through its definition based on $ 3rd $ order cumulants: (1) indirect method and (2) direct method.

\paragraph{Indirect Method}
Let $ x(k),k=1,...,L $ be the available discrete signal. First, we have to segment the data into $ N $ parts each with $ M $ samples. Now $ x_i(k),k=1,...,M $ is the $ i-th $ segment. Then, After subtracting the mean of each segment the $ 3rd $ order moment of each segment is computed as below:
\begin{equation}
m_3^{x_i}(\tau_1,\tau_2)=\frac{1}{M}\sum_{l=l_1}^{l_2}X_i(l)X_i(l+\tau_1)X_i(l+\tau_2)
\end{equation}
where $ l_1=max(0,-\tau_1,-\tau_2) $ and $ l_2=min(M-1,M-2) $.

Since we made each segment zero-mean, its third-order moments and cumulants are equal. Thus, now the average cumulants have to be calculated:
\begin{equation}
c_3^{x}(\tau_1,\tau_2)=\frac{1}{N}\sum_{i=1}^{N}m_3^{x_i}(\tau_1,\tau_2)
\end{equation}

Finally, the third order spectrum (bispectrum) can be estimated as [79]:
\begin{equation}
B_{xxx}(f_1,f_2)=\sum_{\tau_1=-L}^{L}\sum_{\tau_2=-L}^{L}c_3^{x}(\tau_1,\tau_2)\exp(-j(f_1\tau_1+f_2\tau_2))\omega(\tau_1,\tau_2)
\end{equation}
where $ L<M-1 $ and $ \omega(\tau_1,\tau_2) $ is a two-dimensional window, introduced to smooth out edge effects [79]. A complete description of appropriate windows that can be used and their properties can be found in [81]. 

\paragraph{Direct Method}
Regarding to assumed signal $ x(k) $ in previous method, again in this method we have to segment the data and zero-mean each segment similar to previous method. After these steps, the Fourier transform of each segment based on $ M $ points have to be computed:
\begin{equation}
F_{x_i}(k)=\sum_{l=0}^{M-1}x_i(l)\exp(-j\frac{2\pi}{M}lk)
\end{equation}
where $ k=0,1,...,M-1 $. Now the third order spectrum (bispectrum) of each segment is obtained as:
\begin{equation}
C_3^{x_i}(k_1,k_2)=\frac{1}{M}F_{x_i}(k_1)F_{x_i}(k_2)F_{x_i}^*(k_1+k_2)
\end{equation}
which is similar to the first definition of bispectrum. The $ C_3^{x_i}(k_1,k_2) $ need to be computed only in the triangular region $ 0\leq k_2\leq k_1, K-1+k_2<M/2 $. Also a smoothing window for reducing the variance can be performed around each frequency. With these we have:
\begin{equation}
\hat{C}_3^{x_i}(k_1,k_2)=\frac{1}{W^2}\sum_{l_1=-W/2}^{(W/2)-1}\sum_{l_2=-W/2}^{(W/2)-1}C_3^{x_i}(k_1+l_1,k_2+l_2)
\end{equation}
where the smoothing window is of size $ (W \times W) $. Finally, the third order spectrum (bispectrum) is given as [79]:
\begin{equation}
B_{xxx}(f_1,f_2)=\frac{1}{N}\sum_{i=1}^{N}\hat{C}_3^{x_i}(f_1,f_2)
\end{equation}

Both the direct and the indirect methods produce asymptotically unbiased and consistent bispectrum estimates [79].

\section{Major Obstacles}
There are some main obstacles confronting through modern cerebral signal analysis methods, particularly when measuring connectivity in brain. Each of them are introduced below and discussed through next chapter within each represented method.

\subsection{Important Issues Regarding Instantaneous Phase Estimation: Phase Slipping Problem}
Phase Slipping is a notorious problem common in phase estimations from cerebral signals which makes the phase sequence to contain fake (unrelated to brain activity) jumps during low foreground (narrow-band filtered signal) SNRs. This problem is rigorously studied in \cite{ESeraj2, ESeraj1}. In \cite{ESeraj2}, a Monte Carlo based statistical framework is proposed for EEG phase and frequency estimation to overcome the problems associated to previous deterministic approaches (introduced in section 1.1.0.3). Also, in \cite{ESeraj1}, despite the basic illustration of this problem, a straightforward approach for EEG phase estimation based on findings in \cite{ESeraj2} is proposed and is evaluated in a real BCI problem. To help other researches use the findings of these studies, a comprehensive cerebral signal phase analysis toolbox is generated by the authors of \cite{ESeraj2} and \cite{ESeraj1}. Please refer to \cite{ESeraj3} for a detailed description of the provided toolbox

\subsection{Volume Conduction}
One of the most important obstacles confronting in investigating correlation and synchronization between different brain regions using EEG/MEG recordings is a phenomenon known as Volume Conduction (VC). The activity of a particular brain area will be observed in some adjacent electrodes because of conduction in brain volume which produces spurious synchrony. This substantial problem provides false results and shows incorrect related brain regions that can not be neglected. The problem of VC is especially large for scalp EEG and MEG data, because of their low spatial resolution. The difference between estimated connectivity directly from neural sources and the corresponding estimates from scalp recordings is an effect of VC [16].

Several recent methods are presented to overcome this substantial problem and as some of them reported, they obtained successful results. Many of these methods are represented in Section 2.

\subsection{Presence of a Common Reference}
EEG is a bipolar signal recorded from the scalp while MEG is a unipolar record. In a unipolar record, the recording channels are reference free whereas in a bipolar record, channels are constructed by subtracting two unipolar signals captured from two distinct scalp electrodes (i.e. a reference and an arbitrary electrode). 

As reported in [47], the confronting problem here is that bipolar recordings impose a distortion in observed synchrony values, such that in general, they destroy the intended physical interpretation of phase synchrony. It is notable to emphasize that the problem of a common reference is an obstacle when using EEG (and not MEG) recording due to their intrinsic bipolarity. The limitations due to this problem in coherence measurements [48], [49], and phase synchrony [47], are widely investigated previously. It is reported that any contamination in the reference channel will significantly affect the coherence measurements [48], [49]. As mentioned before, one of the superb benefits of using PS over conventional coherence is that PS separates the amplitude information from the phase. Nevertheless, problem of common reference imposes limitations to PS too. It has been shown that if two signals are synchronized, their differences from a third signal are not necessarily synchronized. Based on this, the choice of the reference electrode can considerably affect the synchrony values obtained, up to the point that they almost span the entire interval [0 1] (see [47]).

There are a number of different ways to choose the reference electrode such as \emph{typical EEG reference montages}, which all voltage differences refer to one specific channel, \emph{bipolar montage}, which reference electrode for every channel is different and normally spatially close to it, \emph{average reference} and the \emph{Laplacian reference} which is a mathematical approximation to a reference-free signal. All these techniques are susceptible to common reference problem but it is preferred to use the \emph{Laplacian reference} method [47].

Also as reported in [47], the significance-level in PS measurements will not help and again high synchrony is detected when the reference electrode amplitude is relatively high. As a conclusion, utilizing MEG data in PS measurements and more generally in brain connectivity studies will produce much more reliable results.

\subsection{Noise Sources and Spurious Synchrony}
A noise source or an artifact in cerebral signal is referred to any unwanted contribution to the field potential. This might vary due to different applications. For example, when studying ERPs the spontaneous background EEG/MEG is considered as an artifact while in many other applications the background EEG/MEG is the subject of study. Also there are some common noise sources in EEG recordings that require a preprocessing step where the artifacts are detected and removed. Artifacts generated by the electrical equipment which are near the recording area, contaminations caused bay muscle activity (i.e. blinking, jaw clenching or frowning) and also other contaminant biological signals (i.e. EOG) are such artifacts. Neglecting these kind of noise sources which are uncorrelated to the signals under study in phase synchronization and brain connectivity measurements might impose incorrect PS values and cause \emph{Spurious Synchrony}.

\subsection{Phase Enslaving}
\textquotedblleft Narrow-band filtering tends to transfer phase behavior consistent with the target frequency (or frequency band) from the time region where this holds to adjacent regions where corresponding oscillations are lacking. Conversely, if the signal exhibits distinct oscillations of some particular frequency $ f $, these may strike through to the phase course even if $ f $ is incompatible with the target frequency. Such tends to happen particularly in time-frequency regions where power is low. Since power in EEG signals decays rapidly with increasing frequency, phase enslaving may thus especially affect higher frequency (i.e. gamma) oscillations\textquotedblright [70].

\section{Applications}
Phase analysis has a broad usage in many fields. It has been used widely in communication investigations (i.e. signal phase modulation/demodulation) and also in biomedical researches, particularly in field of neuroscience (i.e. brain functional integration and cerebral signal synchrony measurements).

The most frequent applications in brain connectivity measurements are typically the \emph{Brain-Computer Interface (BCI)}, \emph{Event Related Potential (ERP)} and the \emph{Cognitive} studies. Recent applications include investigation of cognitive processes such as visual perception, mental rotation [50], or pathological states such as epileptic seizures [39], schizophrenia [51], [52], attention deficit (hyperactivity) disorder [53] or migraine [54].

\section{Summary}
In this chapter an overview of frequent cerebral signals, their history and their early and modern analysis methods, particularly in cerebral connectivity measurements were presented.

Most of the early quantities gained by digitalizing the EEG records were introduced. Quantities such as \emph{PSD}, \emph{Correlation Function}, signals' \emph{Instantaneous Phase} and \emph{Amplitude} and also the concept of \emph{Coherence}. In modern analysis the main problem discussed was estimating functional integration between neural regions from spatio-temporal patterns in cerebral signals. To this purpose, the concept of \emph{Synchronization} was introduced. It had been discussed that synchronization can be classified into three main categories: (1) linear, (2) nonlinear and (3) information-based methods. Conventional approaches in these categories were also represented. Finally, the major obstacles confronting in this new fast growing field were discussed and many possible and commonly used applications to this concept were introduced.

The next Chapter is dedicated to delicately representing many introduced \emph{Phase Analysis} methods in various distinct fields which are applicable to brain studies.

\newpage
\chapter{Phase Analysis Methods}
Phase information is generally considered to be \textquotedblleft purer\textquotedblright and less contaminated or more informative than the amplitude of cerebral recordings, which is more influenced by the impedance of the skull or various artifacts such as eyes or face muscles movements [55]. Biological signal phase is an informative useful alternative feature in aforementioned cases that has been of interest in many different research fields through past few years. Thus, using phase of brain signals in cases that amplitude informations are contaminated or inadequate, or in cases that the phase itself has valuable adjunct information over amplitude (i.e. brain cognitive response investigations), would be very helpful.

As mentioned before, measuring brain connectivity is of a particular interest in neuroscience. Because of the reentry mechanism, phase variables and particularly their synchronization for the purpose of measuring connectivity in cerebral signals became more interesting [16].

Various methods within different fields and frameworks are presented in literature for the purpose of cerebral signal phase analysis. The main frameworks and basic concepts were introduced in previous Chapter. In this Chapter, we try to represent and elaborate these various methods within their corresponding field and framework.

\section{Coherence}
\subsection{Conventional MSC}
As introduced earlier in Section (1.1.0.4.1), the conventional method presented in [17] for measuring coherence is based on weighted windowing of the Fourier transform of signals. Let $ x(t) $ and $ y(t) $ be two random process, zero mean, wide-sense stationary and ergodic signals with length $ l $. A summary of the method in [17] is given below:
\begin{itemize}
\item Divide the signals into $ N $ equal parts with $ T $ samples (different part can be overlapped or disjoint)
\item Multiply the samples of each ensemble in a weighting function (i.e. cosine)
\item Take \emph{FFT} of each weighted ensemble
\item Measure power spectral densities of signals (\emph{PSD})
\item Finally, measure the coherence function (or as said \textquotedblleft\emph{Mean Squared Coherence (MSC)\textquotedblright})
\end{itemize}

If $ X_i(f) $ and $ Y_i(f) $ be the Fourier transform of i-th ensemble, the PSDs are calculated as follow:
\begin{equation}
PSD_{xx}(f)=\frac{1}{NT}\sum_{i=1}^{N}|X_i(f)|^2
\end{equation}
\begin{equation}
PSD_{yy}(f)=\frac{1}{NT}\sum_{i=1}^{N}|Y_i(f)|^2
\end{equation}
\begin{equation}
PSD_{xy}(f)=\frac{1}{NT}\sum_{i=1}^{N}X_i(f)Y_i^*(f)
\end{equation}  
where the signals are sampled at $ f_s (f_s>2BW) $ and the $ BW $ is the band width. $ N $ is the number of ensembles with 50\% overlap, $ T $ is the number of samples in ensembles and a $ p-point FFT $ is taken $ (p=Tf_s) $. 

With the ergodicity of signals being neglected we have:
\begin{equation}
PSD_{xx}(f)=E\{|X_i(f)|^2\}
\end{equation}
\begin{equation}
PSD_{yy}(f)=E\{|Y_i(f)|^2\}
\end{equation}
\begin{equation}
PSD_{xy}(f)=E\{X_i(f)Y_i^*(f)\}
\end{equation} 
        
where $ E\{.\} $ illustrates the mathematical expectation. Using these PSDs we compute the mean squared coherence \emph{MSC} as:
\begin{equation}
|MSC(f)|^2=\frac{|PSD_{xy}(f)|^2}{PSD_{xx}(f)PSD_{yy}(f)}=\frac{|\sum_{i=1}^{N}X_i(f)Y_i^*(f)|^2}{\sum_{i=1}^{N}|X_i(f)|^2\sum_{i=1}^{N}|Y_i(f)|^2}
\end{equation}
      
and the complex coherence function is obtained as below:
\begin{equation}
MSC(f)=\frac{PSD_{xy}(f)}{\sqrt{PSD_{xx}(f)PSD_{yy}(f)}}
\end{equation}
         
It is notable that $ T $ have to be chosen large enough to reduce the bias and standard deviation of the measurements. Also the Fourier transform of the weight function have to be narrow. Thus a good resolution for PSD measurements is provided when $ N $ and $ T $ both are chosen large (the best resolution is obtained in 5o\% overlap between ensembles).

The important question confronting here is that \textquotedblleft\textit{what is the optimum way for averaging the MSC over ensembles?}\textquotedblright. As reported in [19], the procedure described above for computing averaged MSC might prevent some coherencies in case of non-stationary phase from being detected. Another method presented for averaging MSC over trials is described below:
\begin{equation}
|MSC(f)|^2=\frac{\frac{1}{N}\sum_{i=1}^{N}|(PSD_{xy}(f))_i|^2}{PSD_{xx}(f)PSD_{yy}(f)}
\end{equation}

\subsection{MSC for Deterministic Evoked Signals}
In some applications, for example detecting brain evoked potentials during a rhythmic periodic stimuli the MSC could be used to measure the correlation and synchronization between evoked potentials and background EEG signal. For Instance, in [56] they assumed $ x(t) $ to be the evoked potential in response to a rhythmic periodic stimuli and $ y(t) $ as the background EEG. If the windowing process is done in the way that all the ensembles contain equal number of the periods of signal $ x(t) $ thus a modification as below is possible. Due to main equation presented for MSC in Section (1.2.1.4) we have:
$$
|MSC(f)|^2=\frac{|\sum_{i=1}^{N}Y_i(f)X_i^*(f)|^2}{\sum_{i=1}^{N}|X_i(f)|^2\sum_{i=1}^{N}|Y_i(f)|^2}
$$

since $ x(t) $ is deterministic, it can be put out from the summation as below:
\begin{equation}
|MSC(f)|^2|_{x(t) periodic}=\frac{|\sum_{i=1}^{N}Y_i(f)|^2}{N\sum_{i=1}^{N}|Y_i(f)|^2}
\end{equation}

where $ 0<MSC(f)<1 $, and it reaches to its maximum value (i.e. 1), when there is only the evoked potentials observing in the windows.

\subsection{Multiple MSC}
The extension of coherence concept and increasing in number of existing signals, particularly in case of multi-channel EEG recordings presents another aspect of coherence known as \emph{multiple coherence}. Multiple coherence shows the relation and synchronization between a signal and a group of other signals. This concept is used to increase the evoked potential detection rate for multi-channel EEG recordings in [57]. Similar to previous method, signal $ x(t) $ assumed to be the brain response of a rhythmic periodic stimuli (i.e. evoked potential) and so deterministic and $ y_i(t)s (i=1,2,...,L)$ as the different \emph{EEG} signals recorded from different electrodes. The coherence function is computed as below:
\begin{equation}
|MSC(f)|^2|_{x:y_1,..y_L}=\frac{PSD_{yx}^H(f)PSD_{yy}^{-1}(f)PSD_{yx}(f)}{PSD_{xx}(f)}
\end{equation}

where $ H $ indicates \emph{Hermitian} of a matrix and the cross-spectrum vector and auto-spectrum matrix are as follow:
\begin{equation}
PSD_{yx}(f)=\left( \begin{array}{c}
PSD_{y_1x}(f)  \\
. \\
. \\
. \\
PSD_{y_Lx}(f) \end{array} \right)
\end{equation}
\begin{equation}
PSD_{yy}(f)=\left( \begin{array}{cccc}
PSD_{y_1y_1}(f) & . & . & PSD_{y_1y_L}(f)  \\
. & . & . & . \\
. & . & . & . \\
PSD_{y_Ly_1}(f) & . & . & PSD_{y_Ly_L}(f) \end{array} \right)
\end{equation}

With $ x(t) $ being deterministic, same as previous method, its Fourier transform has an identical value in all ensembles and a simplification could be made:

\begin{equation}
PSD_{yx}(f)|_{x(t) periodic}=\left( \begin{array}{c}
\sum_{i=1}^{N}Y_{1i}^*(f)X(f)  \\
. \\
. \\
. \\
\sum_{i=1}^{N}Y_{Li}^*(f)X(f) \end{array} \right)
\end{equation}

$$
PSD_{yx}(f)|_{x(t) periodic}=X(f)\left( \begin{array}{c}
\sum_{i=1}^{N}Y_{1i}^*(f)  \\
. \\
. \\
. \\
\sum_{i=1}^{N}Y_{Li}^*(f) \end{array} \right)
$$

\begin{equation}
PSD_{yx}(f)|_{x(t) periodic}=X(f)V(f)
\end{equation}

Putting this result into equation (25) results:
\begin{equation}
|MSC(f)|^2|_{x(t) periodic}=\frac{V^H(f)PSD_{yy}^{-1}(f)V(f)}{N}
\end{equation}

This procedure for measuring MSC is helpful calculating phase synchronization in multi-channel EEG recordings and particularly in application of detecting evoked potentials among a group of other channels of background records. As an example of recent studies using MSC for investigating cognitive sources of specific EEG based features within brain, we can mention \cite{ESeraj6}.

\subsection{MSC Based on Thomson Multitapers}
A novel method based on \emph{Thomson Multitapers} [58], for determining coherence function is presented in [59]. In this method, the left and right singular vectors of the cross-covariance matrix corresponding to a known cross-spectrum model are evaluated as multitapers for estimation of cross-spectrum and coherence. A random process, zero mean, stationary signal $ x(t) $, is filtered through a causal linear filter $ h(n) $ and a stationary sequence $ y(t) $ is created and then the MSC function is measured using a multitaper spectrum estimator as:
\begin{equation}
|MSC(f)^{mt}|^2=\frac{|PSD_{xy}^{mt}(f)|^2}{PSD_{xx}^{mt}(f)PSD_{yy}^{mt}(f)}
\end{equation}

where $ -0.5<f<0.5 $ and:
\begin{equation}
PSD_{xy}^{mt}(f)=\frac{1}{N}\sum_{i=1}^{N}u_i^HF^H(f)XY^HF(f)v_i
\end{equation}
\begin{equation}
PSD_{xx}^{mt}(f)=\frac{1}{N}\sum_{i=1}^{N}u_i^HF^H(f)XX^HF(f)u_i
\end{equation}
\begin{equation}
PSD_{yy}^{mt}(f)=\frac{1}{N}\sum_{i=1}^{N}v_i^HF^H(f)YY^HF(f)v_i
\end{equation}

where $ n $ samples of signals, $ x=[x(0),...,x(n-1)]^T $ and $ y=[y(0),...,y(n-1)]^T $ are used, and $ H $ is showing conjugate transpose and the left and right windows are $ u_i=[u_i(0),...,u_i(n-1)]^T $ and $ v_i=[v_i(0),...,v_i(n-1)]^T $ respectively and $ F=diag[1,e^{-j2\pi f},...,e^{-j2\pi (n-1)f}] $ is the Fourier transform matrix.

According to [59], using this procedure for measuring MSC has some advantages over former methods (i.e. based on weighted overlapped segment averaging) in many applications.

\subsection{Imaginary Part of Coherence}
A new method based on imaginary part of the complex coherence function is presented in [61] and it has been shown that the results of this method can not be caused by \emph{Volume Conduction} (VC). This method assumes that there is no time lag between the potential recorded on the scalp and the source of this potential. This assumption is proved through an investigation in [62]. During an accurate study, they do not observed any time lag or any shift in phase of the recorded potential compared with source of the corresponding signal for frequencies under $ 100Hz $.

A very striking point discussed in [61] is that they argued about using coherency for non-stationary signals and reported that the coherence analysis is feasible in non-stationary processes as strong as stationary ones.

The basic idea used in method presented by [61] is using the real and imaginary part of the coherency instead of its amplitude and phase. Since the real and imaginary parts of coherency are just another representation of complex coherency function, no new quantities are being calculated and it is just a new perspective.

According to [61], the (complex) coherency of the non-interacting areas of brain are necessarily real. Therefore the imaginary part of the coherency can be a good choice for investigation on interacting areas of brain. To see this, assume that the $M$ possible active sources at any time are given as follow:
$$S=[S_{1}, S_{2},...,S_{M}]^{T}$$

and the $L$ scalp electrodes as:
$$X=[X_{1}, X_{2},...,S_{L}]^{T}$$

now we have:
\begin{equation}
\left( \begin{array}{c}
x_{1}(t)  \\
. \\
. \\
. \\
x_{L}(t) \end{array} \right)= \left( \begin{array}{cccc}
a_{11} & . & . & a_{1M} \\
. & . & . & . \\
. & . & . & . \\
a_{L1} & . & . & a_{LM} \end{array} \right) \left( \begin{array}{c}
s_{1}(t)  \\
. \\
. \\
. \\
s_{L}(t) \end{array} \right)
\end{equation}

where $a_{lm}$ denote the degree to which source $s_{m}$ project to electrode $x_{l}$. With the coefficient $a_{lm}$ assume to be constant, we have in frequency domain:
\begin{equation} 
\left( \begin{array}{c}
X_{1}(f)  \\
. \\
. \\
. \\
X_{L}(f) \end{array} \right)= \left( \begin{array}{cccc}
a_{11} & . & . & a_{1M} \\
. & . & . & . \\
. & . & . & . \\
a_{L1} & . & . & a_{LM} \end{array} \right) \left( \begin{array}{c}
S_{1}(f)  \\
. \\
. \\
. \\
S_{L}(f) \end{array} \right)
\end{equation}

Now assume that the signals recorded from two adjacent electrodes, $ x(t) $ and $ y(t) $ are superposed of $ K $ independent sources, then according to above equations we have:
\begin{equation}
X(f)=\sum_{i=1}^{K}a_iS_i(f)
\end{equation}
\begin{equation}
Y(f)=\sum_{j=1}^{K}b_jS_j(f)
\end{equation}

Now based on the main assumption of this method, while computing the cross-spectrum of these signals we will have:
\begin{equation}
PSD_{xy}(f)=E\{X(f)Y^*(f)\}=\sum_{i}\sum_{j}a_ib_jE\{S_i(f)S_j^*(f)\}
\end{equation}
\begin{equation}
\Rightarrow PSD_{xy}(f)=\sum_{i}a_ib_iE\{S_i(f)S_i^*(f)\}=\sum_{i}a_ib_i|S_i(f)|^2
\end{equation}

which is real. Thus the normalization to auto-spectrum terms will also be real and so will be the coherence function. Based on this result and the assumption, if the VC causes no time lag (which has been shown in [62]), the imaginary part of the coherency has no effect of VC artifact in it. This also has been shown in [61] by adding non-interacting signal (i.e. noise) and experiencing a decrease in imaginary part of coherency.

In fact, the imaginary part of coherency shows the phase synchronization between time-lagged signals. This claim is coming from the fact that time-lag in time domain causes a phase shift in frequency domain and vise versa. Therefore, it could be said that the \emph{\textbf{imaginary part of coherency}} specifically shows the synchronization between \emph{\textbf{time-lagged}} signals.

According to [61], however the amplitude and phase of the complex coherence function provide similar information to its real and imaginary part, but looking to the imaginary part instead of phase have some advantages as below:
\begin{itemize}
\item Non-interacting sources do not lead to small but rather to random phases. We cannot interpret a phase without having an estimate of its significance at the same time.
\item One usually calculates coherency with respect to a baseline (a rest condition). Since in the individual coherencies the real parts are typically much larger than the imaginary parts, the phase flips by $\pi$ depending on whether the real part of coherency is larger in the rest or active condition. The interesting structure is easily obscured by this rather meaningless effect.
\item Phase is usually regarded as an additional information about time delay between two processes. However, volume conduction strongly affects the real part but does not create an imaginary part. Processes can appear to be synchronized with almost vanishing time delay while it is only the volume conducted copies of the signals which do not have a time delay.
\end{itemize}

Finally, stochastic given by [61] to measure the imaginary part of coherency is presented in follow. It has been shown that for two signals $ x(t) $ and $ y(t) $, the imaginary part of power spectral density in determined as below:
\begin{equation}
Imag\{X(f)Y^*(f)\}=\sum_{i=1}^{K}\sum_{j=1}^{K}A_iA_j(a_ib_j-a_jb_i)sin(\Delta \theta_{ij})
\end{equation}

where $ i $ and $ j $ indicate the independent sources and finally the imaginary part of coherency (ImagC) is obtained as:
\begin{equation}
ImagC_{xy}(f)=\frac{E\{Imag\{X(f)Y^*(f)\}}{\sqrt{E\{A_x\}^2E\{A_y\}^2}}=\frac{E\{Imag\{PSD_{xy}(f)\}}{\sqrt{E\{A_x\}^2E\{A_y\}^2}}
\end{equation}

which varies between 0 and 1 (similar to $ MSC(f) $).

\subsection{Phase Lag Index}
The ImagC normalizes the components of cross-spectrum to the amplitudes of signals. Thus it is not true if one says it is only related to the imaginary components of cross-spectrum. Adding uncorrelated noise sources causes the amplitude of cross-spectrum to increase while the numerator of ImagC fraction stays unaffected. This problem is investigated in [63] and by simulations it has been shown that the normalization of ImagC to the amplitudes of signals increases its sensitivity to additional uncorrelated noise sources and changes in phase of coherency. Due to this, another method for measuring phase synchronization is represented in [63] to confront these problems. As a potential improvement on the ImagC, [63] proposed the phase lag index (PLI). The PLI estimates, for a particular frequency, to what extent the phase leads and lags between signals from two sensors are non-equiprobable, irrespective of the magnitude of the phase leads and lags. In simulations, the PLI performed better than the ImagC in detecting true changes in phase synchronization, and was less sensitive to the addition of volume conducted noise sources [64]. The main purpose of PLI was to solve two major problems (1) \emph{volume conduction} and (2) \emph{common reference electrode}. 

\textquotedblleft The central idea is to discard phase differences that center around 0 mod $ \pi $. One way to realize this is to define an asymmetry index for the distribution of phase differences, when the distribution is centered around a phase difference of zero. If no phase coupling exists between two time series, then this distribution is expected to be flat. Any deviation from this flat distribution indicates phase synchronization. Asymmetry of the phase difference distribution means that the likelihood that the phase difference $ \Delta \phi $ will be in the interval $-\pi < \Delta \phi < 0$ is different from the likelihood that it will be in the interval $0 < \Delta \phi < \pi$. This asymmetry implies the presence of a consistent, nonzero phase difference (lag) between the two time series. The existence of such a phase difference or time lag, however, cannot be explained by the influences of volume conduction from a single strong source or an active reference, since these influences are effectively instantaneous. The distribution is expected to be symmetric when it is flat (no coupling), or if the median phase difference is equal to or centers around a value of 0 mod $ \pi $\textquotedblright [63]. 

An index of the asymmetry of the phase difference distribution can be obtained from a time series of phase differences $ \Delta\phi_{t_k}, k=1,...,N $ in the following way:
\begin{equation}
PLI= |E\{sign[\Delta\phi_{t_k}]\}|
\end{equation}

Phase lag index measures phase differences as they occur on the unit circle by first thresholding each $ \Delta\phi $ using the signum function and then averaging over successive data points [63]. The PLI ranges between 0 and 1. A PLI of zero indicates either no coupling or coupling with a phase difference centered around 0 mod $ \pi $. A PLI of 1 indicates perfect phase locking at a value of $ \Delta\phi $ different from 0 mod $ \pi $. The stronger this nonzero phase locking is, the larger PLI will be. Note that PLI does no longer indicate, which of the two signals is leading in phase. Whenever needed, however, this information can be easily recovered, for instance, by omitting the absolute value in above equation [63].

\subsection{Weighted Phase Lag Index}
The PLI is said to has some defects. Recently, in a comprehensive study in [64], it has been shown that PLI has some problems with VC, noise sources and detecting changes in phase synchronization due to its discontinuity. It is said that in PLI a small perturbation turns phase lags into phase leads and vise versa. Based on this, another quantity for measuring phase synchronization known as \emph{Weighted Phase Lag Index} (WPLI) is presented in [64]. It is said that WPLI is an improvement over both the conventional PLI and the ImagC. According to [64], the advantage of WPLI over PLI is that it weights the observed phase lags and leads by the magnitude of the imaginary components of cross-spectrum that causes less sensitivity to additional uncorrelated noise sources and more power of detecting phase synchronization statistically.

As mentioned before (see Section 1.4), in case of phase synchronization measurements four problems are well known:
\begin{itemize}
\item The presence of a common reference signal
\item Volume Conduction
\item The presence of noise sources
\item Sample-size bias
\end{itemize}
VC, and in case of EEG (but not MEG) data, the use of a common reference, can spuriously inflate phase synchronization indices. The problem of VS is especially large for scalp EEG and MEG data, because of their low spatial resolution.

Due to these problems, to increase the detection rate of true changes in phase synchronization and also to decrease the effects of additional uncorrelated noise sources, coherency phase changes and the common reference problem, the represented statistics for WPLI are as follow:
\begin{equation}
WPLI(f)=\frac{|E\{Imag\{X(f)Y^*(f)\}\}|}{E\{|Imag\{X(f)Y^*(f)\}|\}}=\frac{|E\{Imag\{PSD_{xy}(f)\}\}|}{E\{|Imag\{PSD_{xy}(f)\}|\}}
\end{equation}
\begin{equation}
WPLI(f)=\frac{|E\{|Imag\{PSD_{xy}(f)\}|sign\{Imag\{PSD_{xy}(f)\}\}|}{E\{|Imag\{PSD_{xy}(f)\}|\}}
\end{equation}

As it is clear, unlike the ImagC, the WPLI uses the components of imaginary part of the cross-spectrum to normalize and also unlike the conventional PLI, $ sign\{Imag\{PSD_{xy}(f)\}\} $ is weighted by $ |Imag\{PSD_{xy}(f)\}| $. The WPLI also varies between 0 to 1 and find its maximum value in:
 $$ sign\{Imag\{PSD_{xy}(f)\}\}=\pm 1 $$
 
  when:
$$ |E\{|Imag\{PSD_{xy}(f)\}|sign\{Imag\{PSD_{xy}(f)\}\}|=E\{|Imag\{PSD_{xy}(f)\}|\} $$

The performance of WPLI is evaluated in confront with the mentioned phase synchronization measurement difficulties in a wide simulation experiment in [64] and it has been reported that the WPLI performed much better in comparison with both PLI and ImagC. As an instance of recent studies using PLI for BCI application we can mention \cite{ESeraj7}.

\subsection{MSC Based on Parametric Spectral Estimation Methods}
Each of these represented method for analyzing of signal phase exhibit some sensitivity to both changes in background power and the instantaneous effects caused by volume conduction. As an improvement to all these methods, the traditional spectral estimation based on Fourier Transform of the raw EEG data, can be replaced by a parametric estimation derived using the \emph{Yule-Walker} AR approach. After this as before, we have to look for periods of phase synchronization [65].

Let $p$ be the model order, $v_i(t)$ be the noise term. For two assumptive signals $ x(t) $ and $ y(t) $ the AR model could be derived as:
\begin{equation}
x(t)=\sum_{q=1}^{p}b_{11}(q)x(t-q)+\sum_{q=1}^{p}b_{12}(q)y(t-q)+v_1(t)
\end{equation}
\begin{equation}
y(t)=\sum_{q=1}^{p}b_{21}(q)x(t-q)+\sum_{q=1}^{p}b_{22}(q)y(t-q)+v_2(t)
\end{equation}

which in frequency domain turns into:
\begin{equation}
X(f)=\sum_{q=1}^{p}B_{11}(q)X(f)e^{-j2\pi fq}+\sum_{q=1}^{p}B_{12}(q)Y(f)e^{-j2\pi fq}+V_1(f)
\end{equation}
\begin{equation}
Y(f)=\sum_{q=1}^{p}B_{21}(q)X(f)e^{-j2\pi fq}+\sum_{q=1}^{p}B_{22}(q)Y(f)e^{-j2\pi fq}+V_2(f)
\end{equation}

To use this model in cross-spectrum measurement and then coherence function estimation using EEG/MEG signals we can extend this to a multi-channel case (i.e. multi-channel EEG). Let $ x(t)=[x_1(t),...,x_L(t)]^T $ be the scalp electrodes. Derivation of \emph{\textbf{Multy-channel AR Model}} gives $ (B(0)=-I , F.T.\{v(t)\}=N(f) $ [65]:
\begin{equation}
x(t)=\sum_{q=1}^{p}b(q)x(t-q)+v(t)
\end{equation}

taking Fourier transform of this equation gives:
\begin{equation}
X(f)=\sum_{q=1}^{p}B(q)X(f)e^{-j2\pi fq}+N(f)
\end{equation}
$$-\sum_{q=1}^{p}B(q)X(f)e^{-j2\pi fq}+X(f)=N(f)$$
$$(I-\sum_{q=1}^{p}B(q)e^{-j2\pi fq})X(f)=N(f)$$
$$(-\sum_{q=0}^{p}B(q)e^{-j2\pi fq})X(f)=N(f)$$
\begin{equation}
\hat{B}(f)X(f)=N(f)
\end{equation}

where $ \hat{B}(f)=-\sum_{q=0}^{p}B(q)e^{-j2\pi fq} $. If $ H(f)=\hat{B}^{-1}(f) $ multiplying in equation above gives the cross-spectrum as:
$$X(f)=H(f)N(f)$$
\begin{equation}
X(f)X^*(f)=H(f)N(f)N^*(f)H^*(f)
\end{equation}

which is an estimation of cross-spectrum needed for establishing the coherence function. Normalization of the temporal averaged version of this cross-spectrum (i.e. expectation) to the amplitude of the estimated  AR coefficients gives the coherence function. To estimate the coefficients of AR model, as presented in [66], the least squares modified Yule–Walker equation (\emph{LSMYWE}) estimator [67], can be used:
\begin{equation}
\hat{b}=-(\hat{R}^H\hat{R}^H)^{-1}\hat{R}^H\hat{r}
\end{equation}
\begin{equation}
\hat{R}=\left( \begin{array}{ccccc}
\hat{r}_{xx,p} & \hat{r}_{xx,p-1} & . & . & \hat{r}_{xx,1} \\
\hat{r}_{xx,p+1} & \hat{r}_{xx,p} & . & . & \hat{r}_{xx,2} \\
. & . & . & . & . \\
. & . & . & . & . \\
\hat{r}_{xx,M-1} & \hat{r}_{xx,M-2} & . & . & \hat{r}_{xx,M-p} \end{array} \right)
\end{equation}
\begin{equation}
\hat{r}=\left( \begin{array}{ccccc}
\hat{r}_{xx,p+1} & \hat{r}_{xx,p+2} & . & . & \hat{r}_{xx,M} \end{array} \right)
\end{equation}

where $ \hat{r}_{xx,n} $ shows the i-th element of auto-correlation estimation for signal $ x_n $, $ p $ is the AR model order and $ M=2p $.

As reported in [65], the parametric and non-parametric spectral estimation methods perform similarly in low noise power while in high noise power the parametric methods act much better in detecting periods of phase synchronization. Using this parametric model in first step of the presented phase synchronization detection methods will improve the results of them and make them less impressible to noise, background power changing and volume conduction effects. The main problem confronting here is dealing with number of channels and the optimum model order.

It is notable that using non-parametric methods have some important advantages too. For instance, parametric methods assume that signals are generated by a linear auto-regressive process but the non-parametric methods do not consider this assumption. Moreover, there are less parameters to be adjusted in non-parametric methods (i.e. the window length), but in parametric methods, after selecting the number of channels and adjusting the model order, the AR model coefficients have to be estimated and after that other parameters such as window length and etc. are set.

\section{Phase Synchronization}
As discussed before, in synchronization concept, coherence deals with several objections. The linearity, stationarity and also the lack of ability to separate the effects of amplitude and phase informations are the main abjections confronting. Thus, another concept called \emph{phase locking} was introduced to overcome these obstacles.

\subsection{Phase Locking Value}
There are two major approaches to calculate this quantity. Both of these methods are based on the framework presented in Section (1.3.2.1). It means in first step they utilize one of the phase extraction methods and then after simply calculating the phase-differences, they quantify the local stability in this phase-differences and finally, in significance-level the significance of captured values are determined. The major difference between methods presented for \emph{Phase Locking Value} (PLV) is in the manner they use to deal with the first and third steps. Due to different techniques utilized in these steps, we represent various statistics for PLV as follow.

\subsubsection{Phase-Locking Statistics}
This method, called \emph{Phase-Locking Statistics} (PLS), measures the significance of the phase-covariance between two signals. PLS separates the phase and amplitude components and can be directly interpreted in the framework of neural integration [34].

To have a meaningful interpretation from instantaneous phase, it has to be extracted through a narrow filter which isolates the frequency of interest. In this method, First of all the signal is band-pass filtered in its frequency of interest and a short arbitrary interval around it (i.e. $ f\pm2 $). After that, its convolution with a complex \emph{\textbf{Gabor-Wavelet}} centered at frequency $ f $ is calculated and phase of this convolution is extracted for all time-bins $ t $, trials $ n=1, . . . , N $, and for each of the pair of electrodes as discussed before in Section (1.2.1.3.2). As mentioned before, when using Wavelet Transform description below gives back the phase-differences which is calculated across trials, as in the second step [34], [35]:
\begin{equation}
\exp(j(\phi_y(t,f)-\phi_x(t,f)))=\frac{W_x(t,f)W_y^*(t,f)}{|W_x(t,f)||W_y(t,f)|}
\end{equation}
where the term $ \phi_y(t,f)-\phi_x(t,f) $ represents the phase-difference between processes $ y(t) $ and $ x(t) $.

As the third step, the stability of phase-differences across trials is quantified by a phase locking value:
\begin{equation}
PLV(t,f)=|\frac{1}{N}\sum_{i=1}^{N}\exp(j(\phi_{y,i}(t,f)-\phi_{x,i}(t,f)))|
\end{equation}
where $ N $ is the total number of trials and $ i $ indicates the index of trial. PLV varies between $ 0 $ and $ 1 $ which indicate the completely non-synchronized and synchronized signals respectively. PLV is said to has phase enslaving problem when the power is low.

As the final step in significance-level, the degree of statistical significance of each PLV value is determined by comparing it to the values obtained by the \emph{surrogate data} which is a statistical test with the null hypothesis that states: \textquotedblleft the value of original data is not significantly different from the specific class of signals\textquotedblright [60]. Efficient surrogate data are generated in a way that they simulate all signals' characteristics, but destroy only the one property suspected to be the source of outstanding values of measure computed [60]. For this purpose, as in [35], the surrogate data is generated by shifting the trials of one of the main signals (i.e. $ y(t) $) in such a way that the phase-differences are no longer computed in simultaneously recorded trials. It means we have:
\begin{equation}
PLV_{surrogate}(t,f)=\frac{1}{K}\sum_{j=1}^{K}|\frac{1}{N}\sum_{i=1}^{N}\exp(j(\phi_{y,j(i)}(t,f)-\phi_{x,i}(t,f)))|
\end{equation}

The proportion of surrogate values greater than the original PLV for a tim $ t $ is called PLS. Normally a criterion of $ 5\% (PLS<5\%)$ is used to characterize significant synchrony [35]. As direct applications of PLV in recent studies we can mention \cite{ESeraj1, ESeraj2, ESeraj5, ESeraj6} and \cite{ESeraj7}. Further implementation details and MATLAB codes related to these studies can be found in \cite{ESeraj3} and \cite{RSameni}.

\subsubsection{Single-trial Phase Locking Statistics}
As an extension of PLS, \emph{Single-trial Phase Locking Statistics} (S-PLS) is presented in [35] in order to estimate PLS in single-trials. S-PLS shares the whole idea with PLS except at a slight cost of temporal resolution, in this method the phase-difference variability is calculated through successive time-steps (not across trials). Based on this, the S-PLV for each individual trial is obtained as below:
\begin{equation}
S-PLV(t,f)=|\frac{1}{\delta}\int_{t-\frac{\delta}{2}}^{t+\frac{\delta}{2}}\exp(j(\phi_{y}(\tau,f)-\phi_{x}(\tau,f)))d\tau|
\end{equation}

As PLV, the S-PLV varies between $ 0 $ and $ 1 $ with $ 1 $ indicating the strongest phase locking [35].

The significance-level in this method is performed similar to that of PLS except that here surrogate values are generated by estimating the maximum PLV between pairs of independent white noise signals. Finally, S-PLS is obtained through a similar criteria to that of PLS.

\subsubsection{Phase Synchrony Based on Shannon Entropy}
The instantaneous phase sequence which is used to estimate the PLV can also be captured through the Hilbert transform (statistics are given in Section 1.2.1.3.1). In a method presented in [36], as the first step they extract the instantaneous phase of signals by Hilbert transform. A merit of the this approach is that the phase can be easily obtained for an arbitrary broad-band signal. Nevertheless, instantaneous amplitude and phase have a clear physical interpretation only if the signals are narrow-band [38]. Therefore, an extra filtering pre-step is required to separate the frequency band of interest. After the pre-step filtering and extracting the phase sequence from the frequency band of interest, Tass et al. [36], proposed that \emph{Shannon Entropy} can be used to characterize the statistical strength of phase synchrony. For this purpose, \textquotedblleft the deviation of the actual distribution of the phase-difference between recording signals from a uniform one must be quantified\textquotedblright [38]. The synchronization index proposed by Tass et al. in [36] can be illustrated as follow:
\begin{equation}
\gamma=\frac{H_{max}-H}{H_{max}}
\end{equation}

where $ H $ is the entropy and defined by (see Section 1.3.3):
\begin{equation}
H=-\sum_{i=1}^{N}P_i\ln P_i
\end{equation}

which N is the total number of trials, and $ H_{max}=\ln(N) $ is the maximal entropy, and $ P_i $ the relative frequency of finding the phase-differences within the $ i-th $ trial [38].

As reported in [36], the optimal number of trials can be obtained as:
\begin{equation}
N=\exp(0.626+0.4\ln(n-1))
\end{equation}

where $ n $ is the number of samples.

Finally, the degree of statistical significance of the phase-locking values was determined through generating surrogate values obtained from surrogate data. Surrogate data is constructed by \emph{scrambling} the original series. This randomization destroys any temporal structure, if present in the original series [36], [38]. The $ \gamma $ is a normalized quantity ($ 0<\gamma<1 $) which measures the strength of phase synchronization and the $ \gamma=1 $ indicates the perfect synchrony.

\subsubsection{Phase Synchrony Based on Mutual Information}
Another phase-locking statistic was introduced in [37]. This method is also similar to previous methods with the whole idea except in procedure presented for quantification the phase synchrony strength. The phase synchrony index in this method is as below:
\begin{equation}
\rho=\frac{I}{I_{max}}
\end{equation}

where $ I_{max=\ln(N)} $ and $ N $ is total number of trials. $ I $ is the mutual information between phase sequences and can be determined as follow:
\begin{equation}
I(\phi_x,\phi_y)=\int_{-\pi}^{\pi}\int_{-\pi}^{\pi}P_{x,y}(\phi_x,\phi_y)\log(\frac{P_{x,y}(\phi_x,\phi_y)}{P_x(\phi_x)P_y(\phi_y)})d\phi_xd\phi_y
\end{equation}

which $ P_x(\phi_x) $ and $ P_x(\phi_y) $ are the probability distribution of the phases $ \phi_x $ and $ \phi_y $, and $ P_{x,y}(\phi_x,\phi_y) $ is their joint distribution. The \emph{Mutual Information} (MI) in this method measures the dependence between phase sequences of signals $ x(t) $ and $ y(t) $.

It has been reported that MI produces surplus amount of information caused by an incorrect assumption that two processes are independent [68]. Also, as discussed in Section (1.3.3), MI can be rewritten as entropy as follow:
\begin{equation}
I(\phi_x,\phi_y)=H(\phi_x)+H(\phi_y)-H(\phi_x,\phi_y)
\end{equation}

Finally, the statistical significant of obtained phase synchrony values is tested through significant-level similar to that of previous method. Normalized in this way, just like previous methods, $ \rho $ varies between $ 0 $ and $ 1 $ where latter indicates the complete synchrony [37].

It seems that this method does not provide any further information in comparison to previous method. Although, more accurate results will be presented in next Chapter after implementing these methods using similar data sets and in a unique situation.

\subsection{Mean Phase Coherence}
Another statistical measure for phase synchrony called as \emph{Mean Phase Coherence} (MPC) is presented in [39]. This method utilize circular variance [69] to characterize the dependences between instantaneous phases of two signals. The method presented in [39] shares the two first steps with previous methods where as first step the instantaneous phases are extracted through Hilbert transform and as second step the phase-differences are calculated by simply subtracting the phase sequences. Finally, as a measure for synchronization the MPC of an angular distribution is defined as:
\begin{equation}
MPC=|\frac{1}{N}\sum_{i=0}^{N-1}\exp(j\phi_{x,y}(i\Delta t))|
\end{equation}

where $ \phi_{x,y} $ indicates the phase differences between phases $ \phi_x $ and $ \phi_y $ and $ \frac{1}{\Delta t} $ is the sampling rate of discrete series. Whit utilizing the Euler’s formula, the above equation turns into:
\begin{equation}
MPC=([\frac{1}{N}\sum_{i=0}^{N-1}\cos(\phi_{x,y}(i\Delta t))]^2+[\frac{1}{N}\sum_{i=0}^{N-1}\sin(\phi_{x,y}(i\Delta t))]^2)^{\frac{1}{2}}
\end{equation}

Clearly from this notation $ MPC $ is restricted to the interval $ [0,1] $ [39]. As same as all previous methods presented for phase synchrony, reliability of these obtained synchrony values is evaluated by the surrogate data approach. In [60] they offered to use \emph{Z-score} statistics for the statistical test in significance-level. Z-score statistic calculates how different is the value of original data from the mean value of all surrogates in the units of standard deviation [60]:
\begin{equation}
Z=\frac{MPC_{org}-\mu(MPC_{surr})}{\sigma(MPC_{surr})}
\end{equation}

where $ \mu $ and $ \sigma $ are the mean value and the standard deviation of all surrogates respectively.

\subsection{Wavelet Coherence}
Another method presented in [35] called as \emph{Wavelet Coherence} (WC) uses the wavelet coefficients of signals and the coherence function to provide a \emph{\textbf{weighted}} version of PLV. The basic idea for presenting this method is that instantaneous phase may be meaningless where power is low and former methods are susceptible to a notorious problems called as \emph{phase enslaving} [70]. AS in [35], WC for time delays $ t $ and frequency(bands) $ f $ can be obtained as below:
\begin{equation}
WC(t,f)=\frac{|\frac{1}{N}\sum_{i=1}^{N}W_{x_i}(t,f)W_{y_i}^*(t,f)|}{\sqrt{\frac{1}{N}\sum_{i=1}^{N}|W_{x_i}(t,f)|^2\frac{1}{N}\sum_{i=1}^{N}|W_{y_i}(t,f)|^2}}
\end{equation}

Since for both signals $ x(t) $ and $ y(t) $ we have:
\begin{equation}
\exp(j\phi(t,f))=\frac{W(t,f)}{|W(t,f)|}
\end{equation}
and also from the original PLV equation:
\begin{equation}
PLV(t,f)=|\frac{1}{N}\sum_{i=1}^{N}\exp(j(\phi_{y,i}(t,f)-\phi_{x,i}(t,f)))|
\end{equation}
then we got:
\begin{equation}
W_{x_i}(t,f)W_{y_i}^*(t,f)=\exp(j\Delta_i(t,f))|W_{x_i}(t,f)||W_{y_i}(t,f)|
\end{equation}
then assuming:
\begin{equation}
\lambda_i(t,f)=\frac{|W_{x_i}(t,f)||W_{y_i}(t,f)|}{\sqrt{\frac{1}{N}\sum_{i=1}^{N}|W_{x_i}(t,f)|^2\frac{1}{N}\sum_{i=1}^{N}|W_{y_i}(t,f)|^2}}
\end{equation}
thus WC can be rewritten as a \emph{weighted} variant of PLV as below [70]:
\begin{equation}
WC(t,f)=|\frac{1}{N}\sum_{i=1}^{N}\exp(j \Delta_i(t,f))\lambda_i(t,f)|
\end{equation}

Due to multiplication by $\lambda_i(t,f)$, wavelet amplitudes are low and this clearly reduces the potential consequences of \emph{phase enslaving}. Nevertheless, WC has some defects such as it does not separate the effects of amplitude and phase in the interrelation between signals [70].

\subsection{Weighted Phase Locking Value}
The basic idea in this method presented in [70] is completely similar to WC method. The only different is between procedures for obtaining the weights. Description below known as \emph{Weighted Phase Locking Value} (WPLV) is presented to measure the statistical strength of phase synchrony between signal pairs:
\begin{equation}
WPLV(t,f)=|\frac{1}{N}\sum_{i=1}^{N}\exp(j \Delta_i(t,f))\omega_{x_i}(t,f)\omega_{y_i}(t,f)|
\end{equation}

In contrast to WC, here weights are not proportional to amplitudes. As offered in [70], the weights differ from $ 1 $ ---which gains the conventional PLV--- only when the amplitudes are considered \textquotedblleft not large enough\textquotedblright and that depends on the respective spectral characteristics. For calculating the weights assume that for signal $ x(t) $ we have (the procedure and statistics for $ y(t) $ is repeated):
\begin{equation}
P_x(t,f)=\frac{1}{f_2-f_1}\int_{f_1}^{f_2}|W_x(t,f)|^2df
\end{equation}
\begin{equation}
P_x(f)=\frac{1}{T}\int_{T}P_x(t,f)dt
\end{equation}
where $ P_x(t,f) $ denotes the mean wavelet power in frequency band $ f $ at time $ t $ and $ P_x(f) $ is its time average across the relevant trial $ T $. Now, assessment of weights can be summarized as below [70]:
\begin{itemize}
\item Large enough: $ \omega_x(t,f)=1 $, if $ P_x(t,f)>\frac{3}{2}P_x(f) $
\item Too small: $ \omega_x(t,f)=0 $, if $ P_x(t,f)\leq\frac{3}{10}P_x(f) $
\item Between: $ \omega_x(t,f)=\frac{\log(\frac{P_x(t,f)}{P_x(f)})-\log(\frac{3}{10})}{\log(5)} $
\end{itemize}

For both the previous and current methods, since the wavelet transform has to be computed anyway, it is preferred to use it also for assessing the instantaneous phases. As discussed before in previous Chapter, the wavelet phases for frequency bands can be determined as the phase angle of the band-averaged complex wavelet coefficients as follow:
\begin{equation}
\exp(j\phi_x(t,f))=\frac{\overline{W}_x(t,f)}{|\overline{W}_x(t,f)|}
\end{equation}
where
\begin{equation}
\overline{W}_x(t,f)=\frac{1}{f_2-f_1}\int_{f_1}^{f_2}W_x(t,f)
\end{equation}

It is notable that \emph{phase enslaving} is less serious problem in case of cerebral signals and it is normally serious when test signals with a sharply defined spectrum is used. Nevertheless, it is expected that WPLV yield more accurate results than the conventional PLV because of discussed problem.

\subsection{Reduced Interference Rihaczek Distribution Phase Synchrony}
This phase synchronization method is based on \emph{Cohen's Class of Distributions} and the \emph{Complex Energy Density Function} (CEDF) phase extraction method. Cohen's class of distributions are bilinear time-frequency distributions and can be illustrated as below [73]:
\begin{equation}
CEDF(t,f)=\int_{-\infty}^{\infty}\int_{-\infty}^{\infty}\phi(\theta,\tau)A(\theta,\tau)\exp(-j(\theta t+2\pi\tau f))d\tau d\theta
\end{equation}
where $ \phi(\theta,\tau) $ is the kernel function and $ A(\theta,\tau) $ is the ambiguity function and can be obtained as below:
\begin{equation}
A(\theta,\tau)=\int_{-\infty}^{\infty}x(u+\frac{\tau}{2})x^*(u-\frac{\tau}{2})\exp(j\theta u)
\end{equation}

\textquotedblleft The major differences between Cohen's class of distributions compared to other time-frequency representations such as the wavelet transform are the nonlinearity of the distribution, energy preservation, and the uniform resolution over entire time-frequency plane\textquotedblright [71]. This distributions describe energy of a signal over time and frequency simultaneously but they can not be used for describing phase information of a signal. Thus, we need complex time-frequency distributions to obtain both  the energy and the phase information. For this purpose, as discussed earlier in Section (1.2.1.3.3), Rihaczek derived the signal energy distribution in time and frequency by application of the complex signal notation. This CEDF, known as Rihaczek distribution, encounters a major disadvantage due to producing crossterms in case of multi-component signals. These crossterms are at the same time and frequency as the original signals and will lead to biased energy and phase estimates [71]. Assume $ z(t)=x(t)+y(t) $, then the Rihaczek distribution is as follows:
\begin{equation}
CEDF(t,f)=\frac{1}{\sqrt{2\pi}}(x(t)X^*(f)\exp(-j2\pi ft)+y(t)Y^*(f)\exp(-j2\pi ft)
$$
$$+x(t)Y^*(f)\exp(-j2\pi ft)+y(t)X^*(f)\exp(-j2\pi ft))
\end{equation}
where the last two terms are crossterms. Therefore, a filtering is needed to remove these crossterms out of calculations.

For this purpose, this method proposes a method based on Cohen's class of distributions and uses its kernel function as the required filter. This is called \emph{Reduced Interference Rihaczek Distribution} (RID) and utilizes the \emph{Choi-Williams} kernel to filter out the crossterms. As in [71], the \emph{Reduced Interference Complex Energy Density Function} or simply (RI-CEDF) can be expressed as:
\begin{equation}
RI-CEDF(t,f)=\int\int\exp(-\frac{(\theta\tau)^2}{\sigma})\exp(j\frac{\theta\tau}{2})A(\theta,\tau)\exp(-j(\theta t+2\pi\tau f))d\tau d\theta
\end{equation}
where $ \exp(j\frac{\theta\tau}{2}) $ is the Rihaczek distribution kernel function and the integrals are from $ -\infty $ to $ -\infty $.

After defining this RI-CEDF, as introduced previously, the phase in time-frequency plane can easily be defined as:
\begin{equation}
\phi_x(t,f)=\arg(\frac{RI-CEDF(t,f)}{|RI-CEDF(t,f)|})
\end{equation}

Now, as the second step in phase synchronization evaluation process, we need to estimate the phase-difference between two signals $ x(t) $ and $ y(t) $. The phase-difference between time-varying phases can be computed as follow:
\begin{equation}
\phi_{x,y}(t,f)=\arg(\frac{RI-CEDF_x(t,f)RI-CEDF_y^*(t,f)}{|RI-CEDF_x(t,f)||RI-CEDF_y(t,f)|})
\end{equation}

Finally in this method, as the final step to assess phase synchrony, the conventional PLV measure is used to quantify the phase synchrony strength between pairs of electrodes.

\subsection{Phase Synchrony Based on Empirical Mode Decomposition}
Another set of methods for evaluating phase synchrony utilize \emph{Empirical Mode Decomposition} (EMD) and separate data into \emph{Intrinsic Mode Functions} (IMFs) and then, extract the instantaneous phases of the IMFs and finally use a phase synchrony quantification procedure, for example in [74], [75] and [77].

Extracting signals phase sequences through EMD algorithm was discussed in previous Chapter. First, signals are decomposed into their relative IMFs and then the Hilbert transform is applied to these IMFs to extract phases. As the next step, for evaluating phase synchrony, the phase-differences have to be calculated. Assuming signals $ x(t) $ and $ y(t) $ and their corresponding phases $ \phi_x(t) $ and $ \phi_y(t) $ which are extracted using EMD, the phase-difference is simply calculated as:
\begin{equation}
\delta\phi_{x,y}(t)=\phi_y(t)-\phi_x(t)
\end{equation}

Finally, as in [74] and [75], the phase synchrony strength is evaluated by this phase-difference through measuring the conventional PLV.

Phase synchrony measurements using EMD have some advantages over other introduced previous methods [74]. First, it is said that the discriminative band of the phase synchrony phenomenon differs not only with subjects, but also with time. Nevertheless, EMD algorithm can decompose the non-stationary signal into its relative IMFs. Therefore, no subject-specific or frequency-specific factors need to be considered to get the discriminative band [74]. Second, band-pass filter is a required necessary step in previous methods which causes time-delay in output signal with respect to the input signal while EMD based methods prevent this time-delay.

As reported in [77], there are two main obstacles confronting while using this method for phase synchrony purposes: (1) it is computationally inefficient (2) misleading results can be obtained as components of different frequencies can temporarily appear synchronized. To overcome these problems, [77] proposes to use the \emph{Complex Extension of EMD}. For more information refer to [77].


\newpage
\appendix
\chapter{Which Coordinates?} \label{App:AppendixA}
since $ PSD_{xy}(f) $ is complex, the \emph{Cartesian} or \emph{polar} coordinates can be used for averaging. In [19] this problem is investigated. Using \emph{Cartesian} coordinate we got:
\begin{equation}
(PSD_{xy}(f))_i=a_i(f)+jb_i(f)
\end{equation}
\begin{equation}
\Rightarrow PSD_{xy}^c(f)=\frac{1}{N}\sum_{i=1}^{N}(a_i(f)+jb_i(f))
\end{equation}

where $ PSD_{xy}^c(f) $ indicates the cross-spectrum averaged in \emph{Cartesian} coordinates and its amplitude and phase are obtaining as follow:
\begin{equation}
|PSD_{xy}^c(f)|=\sqrt{(\frac{1}{N}\sum_{i=1}^{N}a_i(f))^2+(\frac{1}{N}\sum_{i=1}^{N}b_i(f))^2}
\end{equation}
\begin{equation}
\angle PSD_{xy}^c(f)=arctan(\frac{\sum_{i=1}^{N}b_i(f)}{\sum_{i=1}^{N}a_i(f)})
\end{equation}

Similarly, in \emph{Polar} coordinates we have:
\begin{equation}
PSD_{xy}^p(f)=\rho(cos(\phi)+jsin(\phi))
\end{equation}

where $ PSD_{xy}^p(f) $ indicates the cross-spectrum averaged in \emph{Polar} coordinates and $ \rho $ its amplitude and $ \phi $ its phase are obtaining as follow:
\begin{equation}
\rho=\frac{1}{N} \sum_{i=1}^{N}|(PSD_{xy}^p(f))_i|=\frac{1}{N}\sum_{i=1}^{N}\sqrt{a_i^2(f)+b_i^2(f)}
\end{equation}
\begin{equation}
\phi=\frac{1}{N} \angle(PSD_{xy}^p(f))_i=\frac{1}{N}\sum_{i=1}^{N}arctan(\frac{b_i(f)}{a_i(f)})
\end{equation}

It is clear that equations (A.3) and (A.6) and also equations (A.4) and (A.7) are equal when all $ a_i(f)s $ are equal and all $ b_i(f)s $ are equal too. In all any other conditions the amplitude and phase averaged in \emph{Polar} coordinates are larger than the amplitude and phase averaged in \emph{Cartesian} coordinates. Due to EEG signal non-stationarity this condition will not occur and thus as reported in [19], the optimum averaged coherence function performance can be obtained through presentation and averaging the power spectral density functions in \emph{Polar} coordinates.


\end{document}